\documentclass[letterpaper]{mn2e}
\input{psfig}
\def\beq{\begin{equation}}
\def\eeq{\end{equation}}
\def\barray{\begin{eqnarray}}
\def\earray{\end{eqnarray}}

\def\gtsima{$\; \buildrel > \over \sim \;$}
\def\ltsima{$\; \buildrel < \over \sim \;$}
\def\prosima{$\; \buildrel \propto \over \sim \;$}
\def\gsim{\lower.7ex\hbox{\gtsima}}
\def\lsim{\lower.7ex\hbox{\ltsima}}
\def\simgt{\lower.7ex\hbox{\gtsima}}
\def\simlt{\lower.7ex\hbox{\ltsima}}
\def\simpr{\lower.7ex\hbox{\prosima}}
\def\la{\lsim}
\def\ga{\gsim}

\newcommand{\apj}{ApJ}

\newcommand{\mnras}{MNRAS}

\newdimen\hssize
\newdimen\hdsize
\def\omm{\Omega_{\rm m}}
\def\oml{\Omega_{\Lambda}}

\def\m{{\bf m}}
\def\A{{\bf A}}
\def\B{{\bf B}}
\def\ms{m_{\rm star}}
\def\mc{m_{\rm cold}}

\def\mfil{m_{\rm fil}}

\def\mfb{m_{\rm fb}}
\def\dotms{\dot{m}_{\rm star}}
\def\dotmc{\dot{m}_{\rm cold}}

\def\fs{f_{\rm s}}

\def\fsdb07{f_{\rm s,D}}
\def\fc{f_{\rm c}}
\def\frc{f_{\rm rc}}
\def\ffd{f_{\rm d}}

\def\lesssim{\mathrel{\hbox{\rlap{\hbox{\lower4pt\hbox{$\sim$}}}\hbox{$<$}}}}
\def\gtrsim{\mathrel{\hbox{\rlap{\hbox{\lower4pt\hbox{$\sim$}}}\hbox{$>$}}}}

\newcommand{\fof}{{\scshape fof~}}

\newcommand{\es}{\epsilon_{\rm s}}
\usepackage{amsmath}
\newcommand{\z}{\emph{z}}
\newcommand{\comment}[1]{}

\begin{document}


\title[]{
On the puzzling plateau in the specific star formation rate at
$\boldsymbol {z\!=\!2\!-\!7}$
}
\author[S. M. Weinmann, E. Neistein, A. Dekel]
       {\parbox[t]{\textwidth}{
        Simone
        M. Weinmann$^{1}$\thanks{E-mail:weinmann@strw.leidenuniv.nl},
        Eyal Neistein$^{2,3}$, Avishai Dekel$^{4}$
        }\\
\vspace*{3pt}\\
$^1$Leiden Observatory, Leiden University, P.O. Box 9513, 2300 RA Leiden, The Netherlands\\
$^2$Max-Planck-Institute for Extraterrestrial Physics,
Giessenbachstrasse 1, 85748 Garching, Germany\\
$^3$Max-Planck Institut f\"{u}r Astrophysik, Karl-Schwarzschild-Str.1,
Postfach 1317, 85741 Garching, Germany\\
$^4$Racah Institute of Physics, Hebrew University, Jerusalem, 91904, Israel\\
}


\date{}

\pubyear{2010}

\maketitle

\label{firstpage}


\begin{abstract}

The observational  indications for a  constant specific star-formation
rate (sSFR) in the redshift  range $z\!=\!2\!-\!7$ are puzzling in the
context  of current  galaxy-formation models.   Despite  the tentative
nature  of the  data, their  marked conflict  with theory  motivates a
study   of  the   possible   implications.   The   plateau  at   ${\rm
  sSFR}\!\sim\!2$ ${\rm  Gyr}^{-1}$ is  hard to reproduce  because (a)
its level is low compared  to the cosmological specific accretion rate
at $z\!\geq\!6$, (b) it is higher than the latter at $z\!\sim\!2$, (c)
the  natural  correlation  between  SFR  and  stellar  mass  makes  it
difficult to  manipulate their ratio,  and (d) a  low SFR at  high $z$
makes  it hard  to produce  enough massive  galaxies  by $z\!\sim\!2$.
Using a flexible semi-analytic  model, we explore ad-hoc modifications
to  the  standard  physical   recipes  trying  to  obey  the  puzzling
observational  constraints.   Successful  models  involve  non-trivial
modifications,  such  as  (a)  a  suppressed SFR  at  $z\!\geq\!4$  in
galaxies  of   all  masses,  by  enhanced  feedback   or  reduced  SFR
efficiency, following an initial active  phase at $z>7$, (b) a delayed
gas consumption into stars, allowing  the gas that was prohibited from
forming  stars or  ejected at  high $z$  to form  stars later  in more
massive  galaxies, and  (c) enhanced  growth of  massive  galaxies, in
terms  of  either faster  assembly  or  more  efficient starbursts  in
mergers, or by efficient star formation in massive haloes.

\end{abstract}




\section{Introduction}
\label{sec:intro}


The bulk of the stellar mass observed today in galaxies is built up at
high  redshift,  where  star  formation  and mass  assembly  are  very
efficient.  This   makes  observations  at   high  redshift  extremely
important in understanding galaxy  evolution in general.  For example,
the maximum  value of the mean cosmological  star-formation rate (SFR)
density is achieved at redshift 1--3 (e.g. Hopkins \& Beacom 2006).

Although high redshift observations  are naturally plagued with larger
uncertainties than local data, it  is clear that a successful model of
galaxy  formation and  evolution  should match  them within  realistic
error margins. Current models are  often mainly tuned to reproduce the
low redshift  universe, and it  is therefore very instructive  to test
them against high redshift observations.  For example, Fontanot et al.
(2009) and Guo et  al.  (2010) point out that the low  mass end of the
stellar mass function is built too quickly at high redshift in current
semi-analytic models  (SAMs hereafter), while Khochfar  et al.  (2007)
claim that their  SAM can reproduce the evolution of  the faint end of
the luminosity function. It seems however likely that the evolution of
the stellar  mass function  over time alone  is insufficient  to fully
constrain the  models, as was  argued by Neistein \&  Weinmann (2010).
Consequently, it is a crucial next step to compare the observed SFR of
high redshift galaxies to models  in more detail than previously done,
where the main focus was in  trying to match the global star formation
rate density.

Pioneering observational  estimates of the  SFR ($\dot{m}_{\rm star}$)
and  stellar mass  ($m_{\rm  star}$) indicate  that  the specific  SFR
(sSFR, $\dot{m}_{\rm star}/m_{\rm star}$)  is roughly constant in time
throughout the  redshift range $z=2-7$, for galaxies  with roughly the
same  mass $m_{\rm star}  \sim$ $(0.2-1)  \cdot 10^{10}M_\odot$,  at a
level ${\rm  sSFR} \sim  1-2$ ${\rm Gyr}^{-1}$.   (e.g.  Stark  et al.
2009a;  Gonz\'{a}lez et  al.  2010;  Labb\'{e} et  al.  2010a, 2010b).
There  are indications  that this  sSFR plateau  is associated  with a
rather  constant sSFR  within each  galaxy  as it  grows (Papovich  et
al. 2010; Stark et al. 2009a).   The plateau is hard to reconcile with
the current theoretical wisdom for several reasons.

In current models of galaxy formation the high-$z$ SFR is assumed to a
large  extent to  be driven  by  the fresh  gas supply  (see 
Kere\v{s} et al. 2005;
Dekel  et
al. 2009; Bouch\'{e} et al.~2010;  and basically all the SAM; but
see also Narayanan et al. 2010 who discuss a potential merger origin
for the population of submillimeter galaxies at $\z \sim 2$).
The
specific cosmological  accretion rate of baryons  is steeply declining
with time,  $\dot{M}/M \propto  (1+z)^{2.5}$ (Neistein \&  Dekel 2008;
Dekel  et  al. 2009).   In  particular, at  $z  \sim  7$ the  specific
accretion rate is higher than the  observed sSFR by a factor of a few,
and at $z \sim 2$ it is lower than the sSFR by a similar factor.  This
is in marked contrast to the observed plateau in the sSFR.  A constant
sSFR  during the  evolution  of each  galaxy  (main progenitor)  would
require either  an exponential  growth in time  of $m_{\rm  star}$ and
$\dot{m}_{\rm star}$ (this is if most  stars are formed in situ to the
main progenitor),  or a non-trivial  combination of effective  SFR and
stellar  assembly rate as  a function  of time  and mass.   An obvious
related difficulty  is introduced by the  fact that a low  SFR at high
$z$  could  make  it  hard  to  produce  enough  massive  galaxies  by
$z\!\sim\!2$  to match  the bright  end  of the  observed galaxy  mass
function at that epoch.

Given the  marked contrast between  the indicated observation  and the
current  models   of  galaxy  formation,   we  appeal  to   a  special
semi-analytic tool.  Traditional SAMs (e.g Kauffmann et al. 1993; Cole
et al. 2000; De Lucia \& Blaizot 2007) describe the processes that are
responsible for galaxy evolution  by physically motivated recipes that
are  fixed  a priori.  They  are thus  geared  to  solve the  `forward
problem',  i.e., test  to  what  extent the  assumed  set of  physical
recipes  provides a  match to  the observed  properties of  the galaxy
population.   This methodology  is  not ideal  for  exploring a  large
variety  of  physical recipes,  some  of  which  may need  to  deviate
significantly from the standard  assumptions.  Our approach here is to
solve the `inverse problem', where we investigate how the observations
at  high  \emph{z}  constrain  the  basic  recipes,  especially  those
associated  with  the  processes   of  mergers,  star  formation,  and
feedback. This is achievable using  the method of Neistein \& Weinmann
(2010, NW10  hereafter), which allows freedom in  choosing the recipes
of interest,  and a  very efficient exploration  of a  broad parameter
space.

In order  to reproduce the sSFR  plateau, we will try  to either lower
the SFR efficiency after a very  short period of high efficiency at $z
\geq 7$, or  to enhance the suppression of SFR by  feedback at $z \geq
4$, especially in high-mass haloes,  to be followed by an enhanced SFR
in the retained or reincorporated gas at $z \sim 2-3$.  In both cases,
the low  sSFR at high  $z$ makes it  difficult to form  enough massive
galaxies at  $z \sim  1-3$, unless  the rate of  mass assembly  due to
mergers and the associated starbursts are pushed to their limits.

The  models  presented in  this  paper  are  not a  priori  physically
motivated.   However,  we have  tried  to  keep  them simple,  and  to
minimize deviations both from the  standard model and from a monotonic
dependence on  halo mass and time. We  also try to keep  the models as
physically plausible  as possible.   Our aim in  this paper is  not to
find  the  ``right'' model.   The  main  goal  is to  investigate  the
approximate nature of  the needed changes, and how  they impact on the
properties of the galaxy population other than the sSFR.

The measurements  of SFR  and stellar  mass at high  $z$ are  still at
their infancy,  and therefore their interpretation,  in particular the
sSFR plateau, is uncertain  and highly controversial.  The main source
of uncertainty is the obscuration by dust, where several authors bring
convincing arguments for little or no dust at $z \geq 4$ (e.g. Bouwens
et  al. 2009; Finkelstein  et al.  2010), while  others do  apply dust
corrections  and  obtain  higher  values of  sSFR.   
Stellar masses may also be affected by systematic errors. For example,
uncertainties in the treatment of the TP AGB-phase may lead to errors
in the stellar mass estimates at $\z=2-3$ by factors of $\sim$ 2 (e.g. 
Maraston et al. 2006; Magdis et al. 2010a).
 The  disagreement
between  a constant  sSFR and  our  current theoretical  wisdom is  so
pronounced that  it motivates a study of  the theoretical implications
despite the observational uncertainties,  adopting the validity of the
sSFR plateau as a working assumption.  An alternative way out from the
puzzle  might  be to  assume  a  time-dependent  stellar initial  mass
function (IMF) (e.g. Dav\'{e} 2010), but this is kept beyond the scope
of the current paper.

The outline of the paper  is as follows. In section \ref{sec:problem},
we summarize the observed sSFR at high redshift and explain the points
of tension  with theory.   In section \ref{sec:formalism},  we explain
the NW10 method  that we use here.  In  section \ref{sec:explore}, the
main  part of this  paper, we  demonstrate how  a simple  standard SAM
fails to reproduce the sSFR  plateau, and proceed by making controlled
changes to this  model in order to better  reproduce the observations.
In sections \ref{sec:change_quiescent} and \ref{sec:change_mergers} we
discuss possible  physical motivations for these  changes.  In section
\ref{sec:dust} we comment on the observational uncertainties regarding
obscuration  by dust.  In  section \ref{sec:previous_work}  we compare
our    results    to    previous    work.    Finally,    in    section
\ref{sec:conclusions}, we present our conclusions.
Throughout  the paper,  we refer  to  the redshift  range $\z=2-3$  as
``intermediate"  redshift, to  $\z=4-6$  as ``high"  redshift, and  to
$\z>6$ as ``very high" redshift.   Our models are based on dark-matter
merger  trees from  the  Millennium $N$-body  simulation (Springel  et
al.  2005),  and  we  thus  use  WMAP1  cosmological  parameters.   In
particular, we quote masses assuming $h=0.73$.

\section{The problem}
\label{sec:problem}
\subsection{Observations}

\begin{figure}
\centerline{\psfig{figure=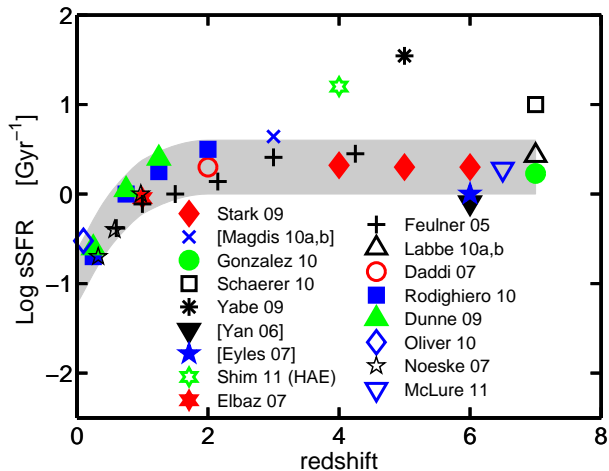,width=3.5in}}
\caption{The observed sSFR plateau.
Shown are measurements of specific star-formation rate as a function of
redshift for galaxies in a similar stellar
mass range $\sim (0.2-1) \cdot 10^{10} M_\odot$.
The references are marked and listed in the text.
The grey belt captures most of the measurements, and reflects
the uncertainty of $\pm 0.3$ dex estimated by Gonz\'{a}lez et al. (2010).
It indicates a constant sSFR $\sim 2$ ${\rm Gyr}^{-1}$ in the range $z = 2-7$,
followed by a steep decline toward $z=0$.
All  references  given in
  \emph{square  brackets}  refer to a rather high median  mass  of $\sim  10^{10}
  M_{\odot}$.
}
\label{fig:obs}
\end{figure}

\A correlation between stellar mass and SFR has
been observed at various redshifts 
up to $z \sim$ 4 (e.g. Daddi et al. 2007; Stark et al. 2009; Karim et al. 
2011).
Figure \ref{fig:obs} shows a compilation of observational estimates
of sSFR as a function of redshift for star-forming galaxies\footnote{typically
selected at high redshift as Lyman-break galaxies (LBGs, Steidel et al. 1999),
which possibly excludes a population of low SFR galaxies
(e.g. Richards et al. 2011). This will however 
only increase differences between
model and basic theoretical predictions.
}
of a similar stellar mass $\sim (0.2-1) \cdot 10^{10} M_{\odot}$.
The data reveal a rather constant sSFR $\sim$ 2 ${\rm Gyr}^{-1}$
in the redshift range $z=2-7$ (Feulner et al. 2005; Yan et al. 2006; Eyles  et al. 2007; Daddi  et al. 2007; Stark  et  al. 2009; Gonz\'{a}lez  et  al. 2010;
Labb\'{e}  et  al. 2010a,b; Rodighiero et  al. 2010;
Gonz\'{a}lez  et  al. 2010; Magdis et al. 2010a,b;
McLure et al. 2011).
except for three
 higher estimates (Yabe et al. 2009; Schaerer \& de Barros
2010, Shim et al. 2011)\footnote{We obtained part of the estimates by dividing the median
SFR by the median stellar mass (for Yan et al. 2006; Eyles et al. 2007;
Yabe et al. 2009). In some other cases, the estimates are based on an
extrapolation of the sSFR-stellar mass relation to a stellar mass
of $\sim 0.5 \cdot 10^{10} M_{\odot}$ (Daddi et al. 2007; Rodighiero et al.
2010).}.
The main reason for these higher estimates is the larger correction
for dust extinction assumed by these authors (to be discussed in
section \ref{sec:dust}), as well as different treatments of nebular emission
lines and different assumptions concerning star-formation histories
(to be discussed in section \ref{sec:individual}). Additionally,
Shim et al. (2011) only include galaxies with indications for H$\alpha$
emission, which will
bias the estimate of the sSFR high. We note that
all the estimates above do not include submillimeter galaxies, which are outliers
to the relation between stellar mass and SFR, simply because these tend
to have stellar
masses above the limit we consider here (e.g. Daddi et al. 2007).
At $z<2$, the
sSFR declines steeply (Noeske et al. 2007; Elbaz et al. 2007; Dunne et al. 2009; Oliver et al. 2010; 
Rodighiero et al. 2010).
The grey belt in Fig. \ref{fig:obs} tries to capture the overall trend,
reflecting an uncertainty of $\pm 0.3$ dex as estimated by
Gonz\'{a}lez et al. (2010), and ignoring the two high estimates.
As will be discussed below, this observed sSFR plateau is puzzling ---
its level is surprisingly high at $z\sim 2$ and surprisingly low at $z >4$.
For the purpose of the theoretical analysis of the current paper,
we adopt the sSFR plateau as marked by the grey belt.

\subsection{Tension with theory}

Here,  we outline  the main  potential points  of tension  between the
observed  sSFR  and   theoretical  predictions  both  from  relatively
detailed SAMs and simple analytical arguments.

\subsubsection{Tension with SAMs}

\begin{figure}
\centerline{\psfig{figure=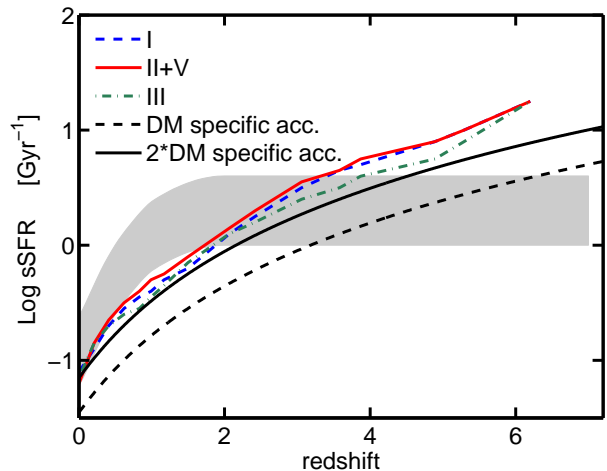,width=3.5in}}
\caption{
Evolution of sSFR in the SAMs of NW10, 
and of the specific dark matter accretion rate.
Shown are four different models (I, II, III and V, with results
of models II and V being indistinguishable and thus represented
by one line) by the SAM of NW10
for galaxies in the mass range  [$2\cdot10^9$,  $10^{10}$] $M_\odot$
(curves in colour).
The completeness limits trying to mimic the observed ones are described in
the text.
In all models the sSFR is steeply declining in time, not reproducing the
observed sSFR plateau marked by the grey belt from Fig.~1.
Also shown is the specific dark matter
 accretion rate onto haloes
of log$(M_{\rm halo}) \sim 10^{12} M_{\odot}$
according to Neistein \& Dekel (2008) (dashed black line),
and the same quantity multiplied by a factor of 2, to account for the effect
of instantaneous mass loss from newly formed stars, as assumed in the
models (solid black line).
}
\label{fig:nw10_ssfr}
\end{figure}

\begin{figure}
\centerline{\psfig{figure=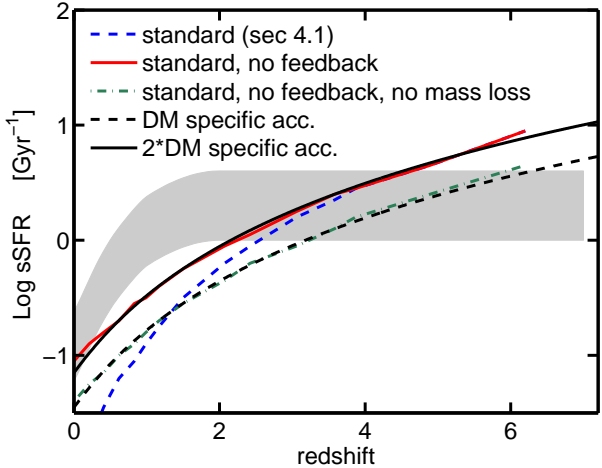,width=3.5in}}
\caption{Evolution of sSFR in variations
of the standard SAM used in this work, and of the 
specific dark matter accretion rate.
The predicted dark matter accretion rate onto haloes
with log$(M_{\rm halo}) \sim 10^{12} M_{\odot}$
according to Neistein \& Dekel (2008) (dashed black line) and multiplied
by a factor of 2 (black line), compared to the standard
model (blue dashed line,
described in section \ref{sec:standard}), the standard model
without feedback (red solid line), and the standard model
without feedback and stellar mass loss
(green dot-dashed line).
The sSFR in the raw model without feedback and mass loss matches the total
specific accretion rate. The mass loss adds a factor of two.
}
\label{fig:DMrate}
\end{figure}

In Fig.~\ref{fig:nw10_ssfr},  we show the sSFR as a function of redshift
at a fixed mass for four of the models presented in NW10, in comparison
with the
observed sSFR plateau.
To  account  for the  observational
completeness limit as  indicated by Stark et al.  (2009), we only take
into  account model  galaxies with  log(SFR)  $>$ 0.25,  0.45 and  0.5
$\rm{yr}^{-1}$  at z=4,  5 and  6 respectively.   The
four models are
described in detail in NW10 with the same numbering used here
 and can be summarized as follows: I) model
without  SN feedback,  II) model  without ejective  SN  feedback, III)
model  including cold  accretion [which is the  model most  similar to
other standard SAMs],  and
V)
 model in
which  cooling and  star  formation shuts  down  after major  mergers.
All of those models have been tuned to reproduce key observables like
the stellar mass functions at different redshifts, and star formation
rate at z=0.
Remarkably, all  models show an  extremely similar
behaviour despite  their fundamental differences,  all in disagreement
with observations.  They overestimate  the sSFR by  about an  order of
magnitude at $z \sim 6$, and  underestimate it by around 0.3 dex at $z
\sim 2$.

Other current SAMs show a similar behaviour.  For  example, Lacey et al. (2010) show that the
sSFR  of  galaxies  in  the  Baugh  et al.  (2005)  model  have
sSFR $>10$ ${\rm Gyr}^{-1}$ at z=6, an order of magnitude higher
than observational results. Daddi et al. (2007)
indicate that their observed SFR at $z \sim 2$ is significantly higher
than the values predicted by the model of Kitzbichler \& White (2007).
Finally, we have confirmed ourselves that the model of
De Lucia \& Blaizot (2007) predicts results very similar to our model
predictions shown in Fig.~2 (see also Guo \& White 2008, their Fig. 3).

\subsubsection{Tension with basic theoretical considerations}
\label{sec:tension1}

The observed sSFR plateau is in disagreement with the standard
wisdom concerning galaxy evolution.
First, the average specific
accretion rate into dark-matter haloes of a given
mass is rapidly increasing  with  redshift, roughly in proportion to
$(1+z)^{2.5}$  (Neistein \&  Dekel 2008).
Second, galaxies are more  dense and gas rich at high-redshift,
which is expected to lead to higher SFR (e.g. Dutton et al. 2010).  In
agreement with these studies, Bouch\'{e} et al. (2010) find
in their idealized model that the sSFR of
individual galaxies is indeed monotonically decreasing with time.
To illustrate the first point, we show
in Fig. \ref{fig:nw10_ssfr} the approximation for the
average specific dark matter accretion rate $\dot{M}/M$
for haloes with log$(M_{\rm halo}) \sim 10^{12} M_{\odot}$
(Neistein \& Dekel 2008), and compare it to the standard model,
described in section \ref{sec:standard}, in Fig. \ref{fig:DMrate}.
Remarkably, when we remove feedback from the standard model,
it
predicts a sSFR evolution in excellent agreement with twice
the specific dark matter accretion rate. As shown in the figure,
the remaining factor of two difference
is fully explained by instantaneous
stellar mass loss.
Given that the ratio between stellar mass
and halo mass is only about 5\%
for the galaxies
considered here, this agreement is noteworthy.
We see in the figure that the feedback as implemented in the standard
model does not have a significant effect on the sSFR at $z>4$, while
it gradually reduces the sSFR at lower redshifts. Compared
to the models of NW10 shown in Fig. \ref{fig:nw10_ssfr}, the standard
model that we use in this paper
has a low sSFR at $z<2$, due to efficient feedback at late times.
This is not relevant for studying the plateau.

In what follows, we will demonstrate that despite this
serious tension with theory, the sSFR plateau can in
principle be reproduced by models of galaxy evolution, but it takes
non-negligible modifications to common ingredients of these models.

\section{The Formalism}
\label{sec:formalism}

In  this section we  describe the  formalism we  use for  modeling the
evolution  of galaxies.  For more  details about  the  methodology the
reader is referred to NW10. It  is shown there that the results of our
model are very similar to those  given by a standard SAM, although the
recipes are simplified and schematic. In the context of this work, the
simplicity of  the model allows us  to tune it  easily, without losing
the  complex  interplay  between  different process  like  dark-matter
growth, cooling, SF, feedback, and  merging. The code is available for
public        usage       through       the        Internet       (see
\texttt{http://www.mpa-garching.mpg.de/galform/sesam})

\subsection{Merger trees}

We use merger trees  extracted from the Millennium $N$-body simulation
(Springel et al. 2005). This simulation was run using the cosmological
parameters   $(\omm,\,\oml,\,h,\,\sigma_8)=(0.25,\,0.75,\,0.73,\,0.9)$,
with a particle mass of $8.6\cdot 10^8 h^{-1}M_{\odot}$ and a box size
of  500  $h^{-1}$Mpc.   The  merger  trees  used  here  are  based  on
\emph{subhaloes}  identified   using  the  \textsc{subfind}  algorithm
(Springel et  al. 2001). They are  defined as the  bound density peaks
inside  \fof  groups  (Davis  et  al.  1985).   More  details  on  the
simulation and  the subhalo merger-trees  can be found in  Springel et
al.  (2005)  and  Croton
 et  al.  (2006).  The  mass of  each  subhalo
(referred to as  $M_{\rm h}$ in what follows)  is determined according
to the  number of particles it  contains.  Within each  \fof group the
most  massive subhalo  is termed  the central  subhalo of  this group.
Throughout this paper we will use the term `haloes' for both subhaloes
and the central (sub)halo of \fof groups.

\subsection{Quiescent evolution}
\label{sec:model_quiescent}

Each galaxy is modeled by a 4-component vector, %
\begin{eqnarray}
\m = \left(  \begin{array}{c} \ms \\ \mc \\  \mfil \\ \mfb \end{array}
\right) \, ,
\label{eq:m_vec}
\end{eqnarray}
where $\ms$  is the  mass of stars,  $\mc$ is  the mass of  cold gas
within the disk,  $\mfil$ is the mass within  cold filaments streaming
within the host  halo into the central galaxy, and  $\mfb$ is the mass
currently made unavailable for  star formation by stellar feedback. We
use  the  term `quiescent  evolution'  to  mark  all the  evolutionary
processes of a galaxy, except those related to mergers.

All the models in this work assume that fresh gas is added
to  a  galaxy   only  by  cold  filaments,  increasing   the  mass  of
$\mfil$. The infall rate into
 filaments is assumed to be proportional to
the dark-matter growth rate, %
\begin{equation}
\left[\dot{m}_{\rm fil}\right]_{\rm accretion}  = 0.17 \, \dot{M}_{\rm
  h} \\
\label{eq:smooth_acc}
\end{equation}
Here 0.17 is the cosmic  baryonic fraction, and $\dot{M}_{\rm h}$ is
the  rate  of dark-matter  smooth  accretion  which  does not  include
mergers with resolved progenitors (if $\dot{M}_{\rm h}<0$ we use a gas
accretion rate of zero).

The mass of cold gas within the disk is increased due to the
free  infall  of cold  filaments  from the  outer  parts  of the  host
halo. We mimic this effect by  assuming that gas joins the disk with a
specific rate $\fc$,  %
{\begin{equation} \left[\dotmc\right]_{\rm ff}
= -\left[\dot{m}_{\rm fil} \right]_{\rm ff} = \fc \cdot \mfil \,.
\end{equation}
The  efficiency $\fc=\fc(M_{\rm h},t)$  is a  function of  the host
halo mass  $M_{\rm h}$ and the cosmic  time $t$ only, and  is given in
units of Gyr$^{-1}$.

We assume that the SF rate is proportional to the amount of cold
gas, %
\begin{equation}
\left[\dotms\right]_{\rm   SF}  =   -\left[\dotmc\right]_{\rm   SF}  =
\fs\cdot \mc \,,
\label{eq:sf}
\end{equation}
where  $\fs=\fs(M_{\rm h},t)$  is a  function of  the halo  mass and
time, in  units of Gyr$^{-1}$.  For  each SF episode we  assume that a
constant  fraction  of the  mass  is returned  back  to  the cold  gas
component  due to  SN  events  and stellar  winds.  This recycling  is
assumed to be instantaneous, and contributes %
\begin{equation}
\left[\dotmc\right]_{\rm    recycling}   =   -\left[\dotms\right]_{\rm
  recycling} = R \left[\dotms\right]_{\rm SF} \,.
\end{equation}
Following NW10, we use $R=0.5$ for all models. This
is the recycled fraction for a Chabrier (2003) IMF at 13.5 Gyr after a star burst
according to the Bruzual \& Charlot (2003) stellar population models. We note that
this is the only point where the assumption on the IMF enters our model.

Cold gas  can be  affected by  feedback, which
means that  it becomes unavailable  for star formation and  moves from
the cold phase  to the feedback phase. Assuming  that stellar feedback
immediately  follows  star  formation,   this  feedback  should  be  in
proportion to the SF rate, %
\begin{eqnarray}
\lefteqn{     \left[\dot{m}_{\rm    fb}\right]_{\rm     feedback}    =
  -\left[\dotmc\right]_{\rm   feedback}   =  }   \\   \nonumber  &   &
\ffd\left[\dotms\right]_{\rm SF} = \ffd \fs \mc \,.
\end{eqnarray}
We model feedback
  by  a function  of  halo  mass and  time,
$\ffd=\ffd(M_{\rm h},t)$.

Once  gas  has  been  made  unavailable  for  star
formation due  to feedback, we allow  it to return to  the cold phase,
with a re-incorporation efficiency: %
\begin{eqnarray}
\left[\dotmc\right]_{\rm rc} = -\left[\dot{m}_{\rm fb}\right]_{\rm rc}
= \frc m_{ \rm fb}.
\end{eqnarray}
Note that the feedback mechanism we  use is moving gas from the cold
phase  to  the  feedback  phase.  The mass  within  filaments  is  not
participating in the feedback and re-incorporation processes.
If the cold gas mass becomes negative in a given timestep due to
strong feedback, the star formation rates are adjusted such that
a cold gas mass of zero is produced.

To conclude, each  process is described by one  function which depends
on the host  halo mass and time only. All  processes discussed in this
section  can be  written  in a  compact  form by  using the  following
differential equations: %
\begin{equation}
\dot{\m} = \A\m + \B\dot{M}_{\rm h} \, ,
\label{eq:m_evolve}
\end{equation}
where
\begin{eqnarray}
\A = \left( \begin{array}{cccc} 0 & (1-R)\fs  & 0 & 0 \\ 0 & -(1-R)\fs
  -\ffd \fs &  \fc & \frc \\ 0  & 0 & -\fc &  0 \\ 0 & \ffd \fs  & 0 &
  -\frc
\end{array} \right)
\end{eqnarray}
\begin{eqnarray}
\B = \left( \begin{array}{c} 0 \\ 0 \\ 0.17 \\ 0
\end{array} \right)  \,\, .
\label{eq:AB_defs}
\end{eqnarray}

Photoionization  heating of  the  intergalactic medium  is assumed  to
suppress  the amount  of cold  gas available  for SF  within  low mass
haloes. This  effect is critical  for modeling the formation  of dwarf
galaxies.   The   minimum   halo   mass  of   $\sim   2\cdot   10^{10}
h^{-1}M_{\odot}$ in the Millennium simulation, which we use here, does 
however
not allow a detailed modeling of small mass galaxies.
Thus, instead of implementing a detailed treamtment of reionization, 
we simply assume that all the gas is kept hot until redshift 9, 
where cooling and SF are allowed to start. This is a
higher redshift than in NW10, which
we found is needed
to  produce a  high enough number  of galaxies at  very high
redshifts.

\subsection{Mergers and satellite galaxies}
\label{sec:model_mergers}

Satellite galaxies  are defined  as all galaxies  inside a  \fof group
except   the   main  galaxy   inside   the   central  (most   massive)
subhalo. Once  the subhalo corresponding to a  given galaxy cannot
be  resolved anymore,  it  is  considered as  having  merged with  the
central halo. Due  to the effect of dynamical  friction, the galaxy is
then assumed to spiral towards the  center of the \fof group and merge
with the galaxy in the central halo after a significant delay time.

At the  last time the dark  matter subhalo of a  satellite galaxy is
resolved  we  compute its  distance  from  the  central halo  ($r_{\rm
  sat}$), and  estimate the dynamical friction time  using the formula
of Binney (1987), %
\begin{equation}
t_{\rm  df}   =  \alpha_{\rm  df}  \cdot  \,   \frac{1.17  V_v  r_{\rm
    sat}^{2}}{G m_{\rm sat}\ln\left(  1+ M_{\rm h}/m_{\rm sat} \right)
} \, .
\label{eq:t_df}
\end{equation}
For $m_{\rm sat}$ we use the baryonic (stars + cold gas) mass of the
satellite galaxy plus  the minimum subhalo mass which  can be resolved
by  the  Millennium  simulation.   $V_v,\,M_{\rm h}$  are  the  virial
velocity and mass of the central subhalo.  If a satellite falls into a
larger halo  together with its  central galaxy we update  $t_{\rm df}$
for both objects according to the new central galaxy.

While  satellite galaxies move  within their  \fof group,  they suffer
from loss of their extended gas reservoir
 due to tidal  stripping. We assume  that all satellite
galaxies are losing their  reservoir of filament gas exponentially, on
a time scale of a few  Gyr.  In order to properly model this stripping
we modify  $\A$ by subtracting a  constant $\alpha_h$ from  one of its
elements:
\begin{eqnarray}
\A_{\rm sat}(3,3) = -\fc-\alpha_h \,.
\label{eq:ABsat_defs}
\end{eqnarray}
Note that  a constant in the  diagonal of $\A$  gives an exponential
time dependence. However, the actual dependence of $\mfil$ on time for
satellite  galaxies  is more  complicated  due  to contributions  from
accretion.  In  general the parameter $\alpha_h$ should  depend on the
dynamical time of the host halo. For simplicity we consider it to be a
constant here. We assume that the gas which is in the feedback
phase is not stripped.

When galaxies finally merge we assume that a SF burst is triggered. We
follow  Mihos et  al.  (1994), Somerville  et  al. (2001)  and Cox  et
al. (2008) and model the amount of stars produced by %

\begin{equation}
\Delta \ms = f_{\rm burst} (m_{1,{\rm cold}}+m_{2,{\rm cold}}) \, ,
\label{eq:sf_burst}
\end{equation}
where
\begin{equation}
f_{\rm burst} =  \alpha_b \left( \frac{m_1}{m_2} \right)^{\alpha_c} \,
.
\end{equation}

Here $m_i$  are the baryonic  masses of the progenitor  galaxies (cold
gas  plus stars),  $m_{i,{\rm  cold}}$  is their  cold  gas mass,  and
$\alpha_b$, $\alpha_c$ are constants.  %

The  burst   duration  has  been  shown  to   vary  in  hydrodynamical
simulations between tens  of Myr to a few Gyr  depending on the merger
mass ratio,  and whether  multiple bursts are  considered or  just the
main peak (Cox et  al. 2008).  We use a timescale of  10 Myr in all our 
models, following
De  Lucia   \&  Blaizot  (2007).
Merger-induced bursts cause feedback in the same way as quiescent star
formation.

\section{Modifying a simple model}
\label{sec:explore}

Below, we  present a simple standard  SAM, which we use  as a starting
point for  our tuning procedure.  This simple model  includes features
common  to many  current models  of galaxy  formation. It  is  kept as
simple  as  possible  in  order  to  facilitate  tuning and
to simplify interpretation of changes to the model.   In  section
\ref{sec:tuning}, we present 6  alternative models which reproduce the
sSFR plateau.

\begin{figure}
\psfig{file=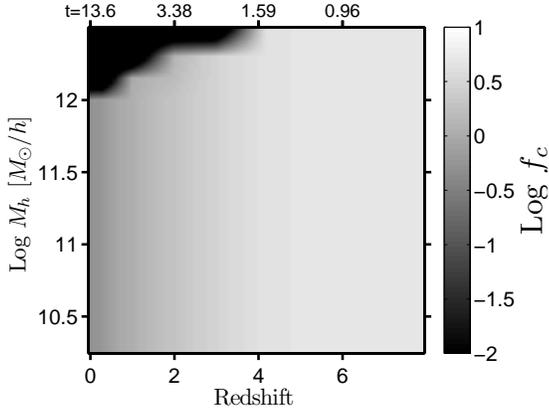,width=220pt}
\caption{The cooling efficiency as a function of halo mass and redshift,
as used in all the models described in sec. \ref{sec:explore}.
}
\label{fig:cooling}
\end{figure}

\begin{figure}
\begin{tabular}{cc}
\psfig{file=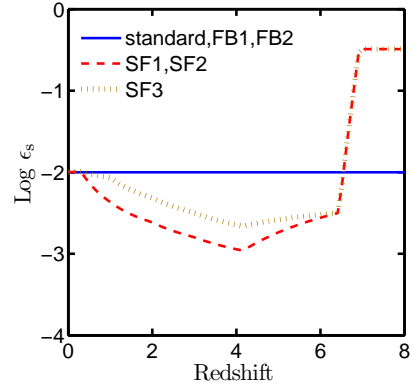,width=220pt}\\ \psfig{file=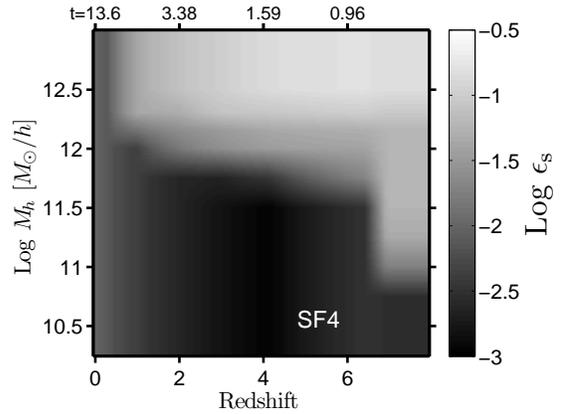,width=220pt}\\
\end{tabular}
\caption{Star formation efficiency, $\es$, as a function of redshift and halo
mass, in the different models as indicated.
Top: in the standard model and in models FB1 and FB2, $\es$ is constant
at all times and halo masses. In models SF1, SF2, and SF3, 
it is a function of redshift only.
Bottom: in model SF4, $\es$ is a function of both redshift and halo mass.}
\label{fig:sf_eff}
\end{figure}

\begin{figure*}
\begin{tabular}{cc}
\psfig{file=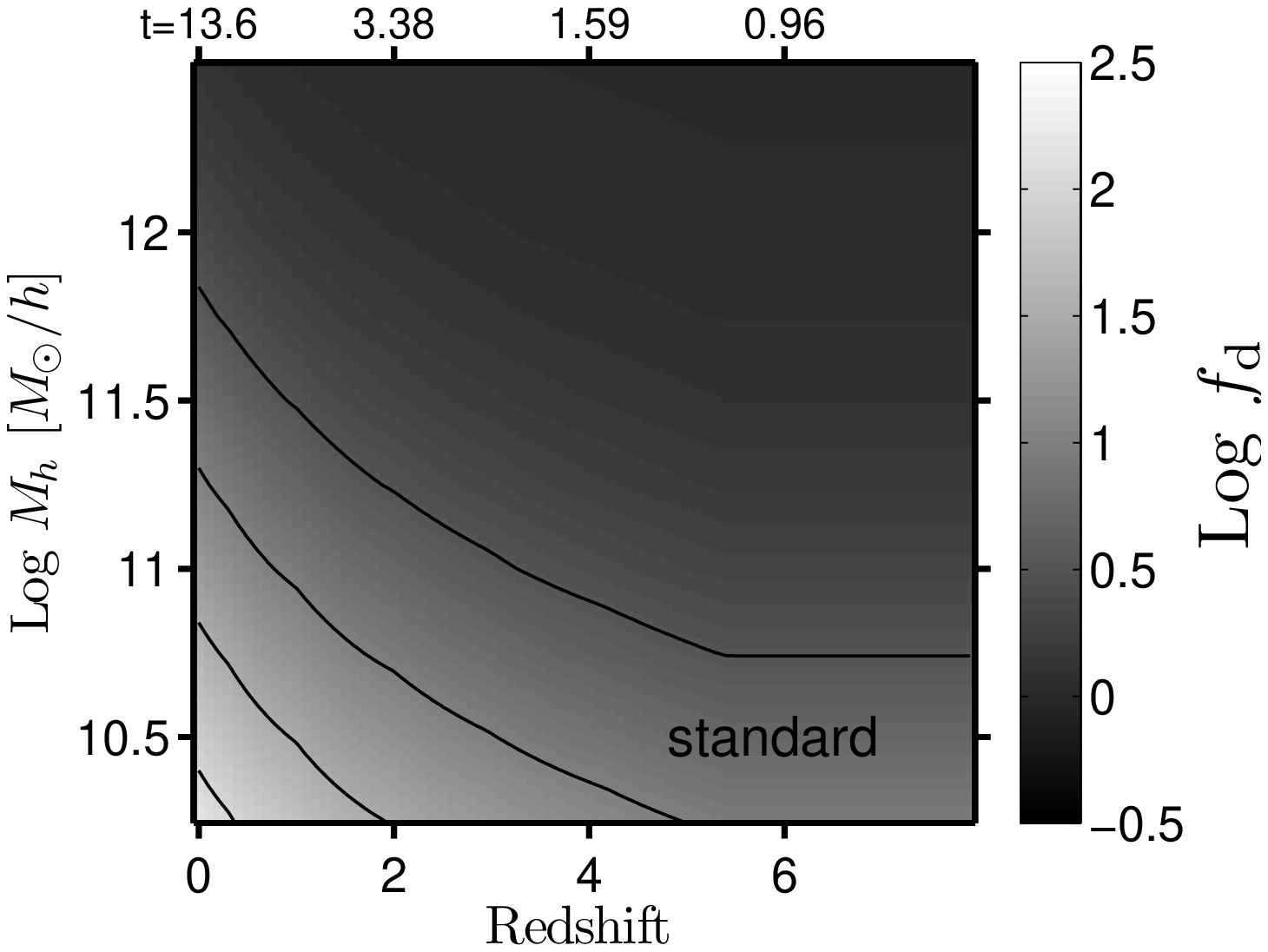,width=220pt}                              &
\psfig{file=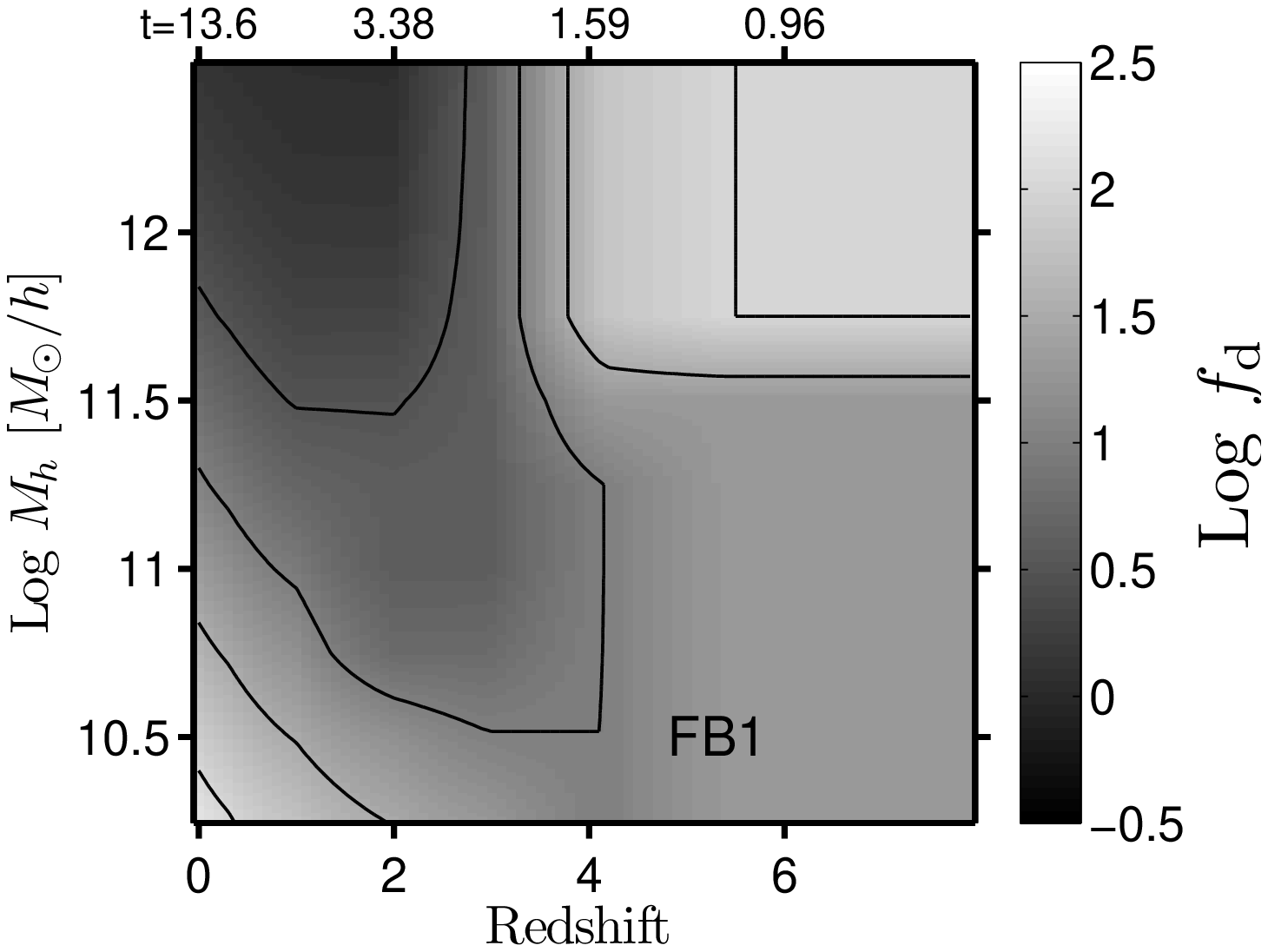,width=220pt}\\ \psfig{file=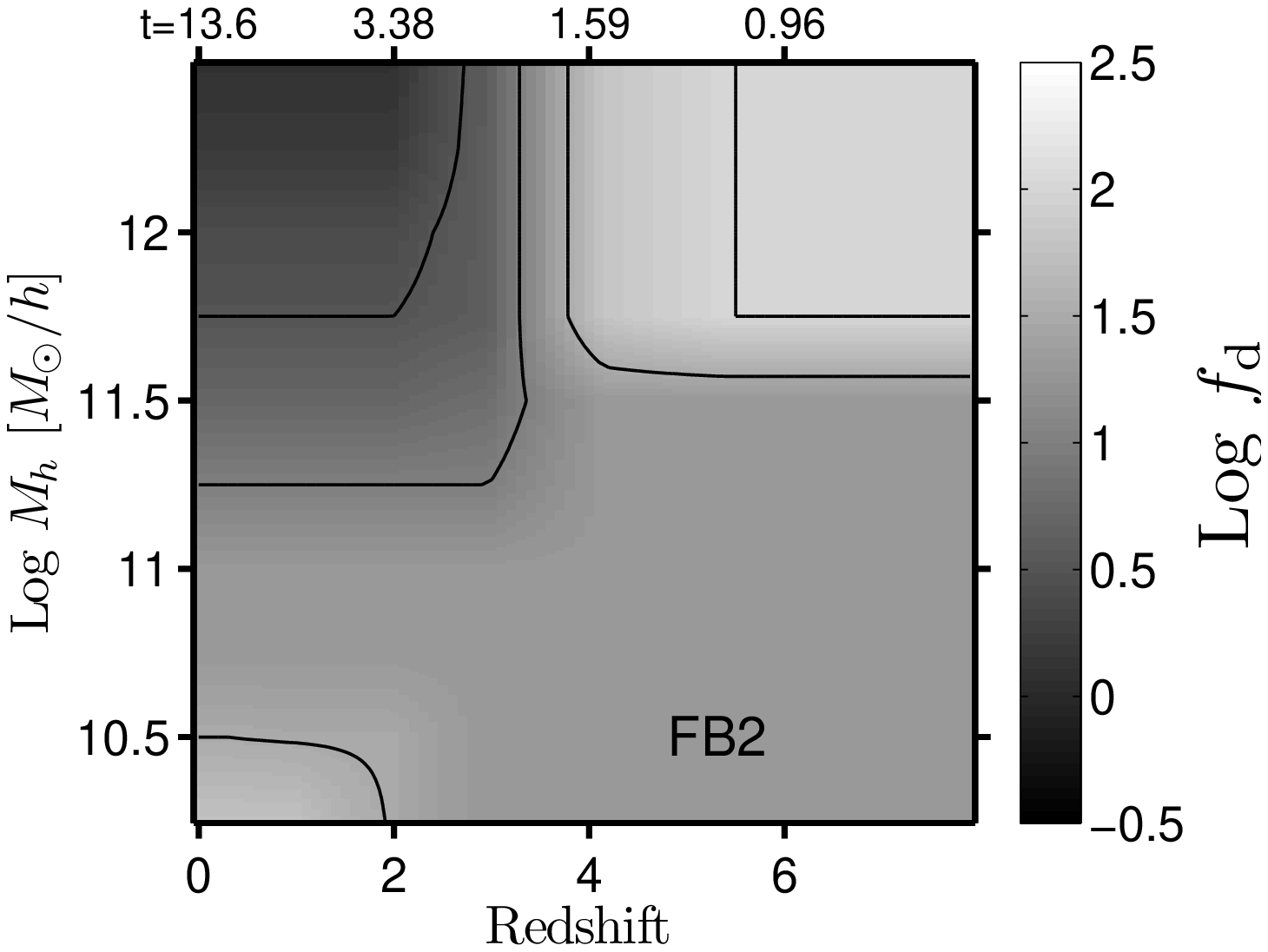,width=220pt}&
\psfig{file=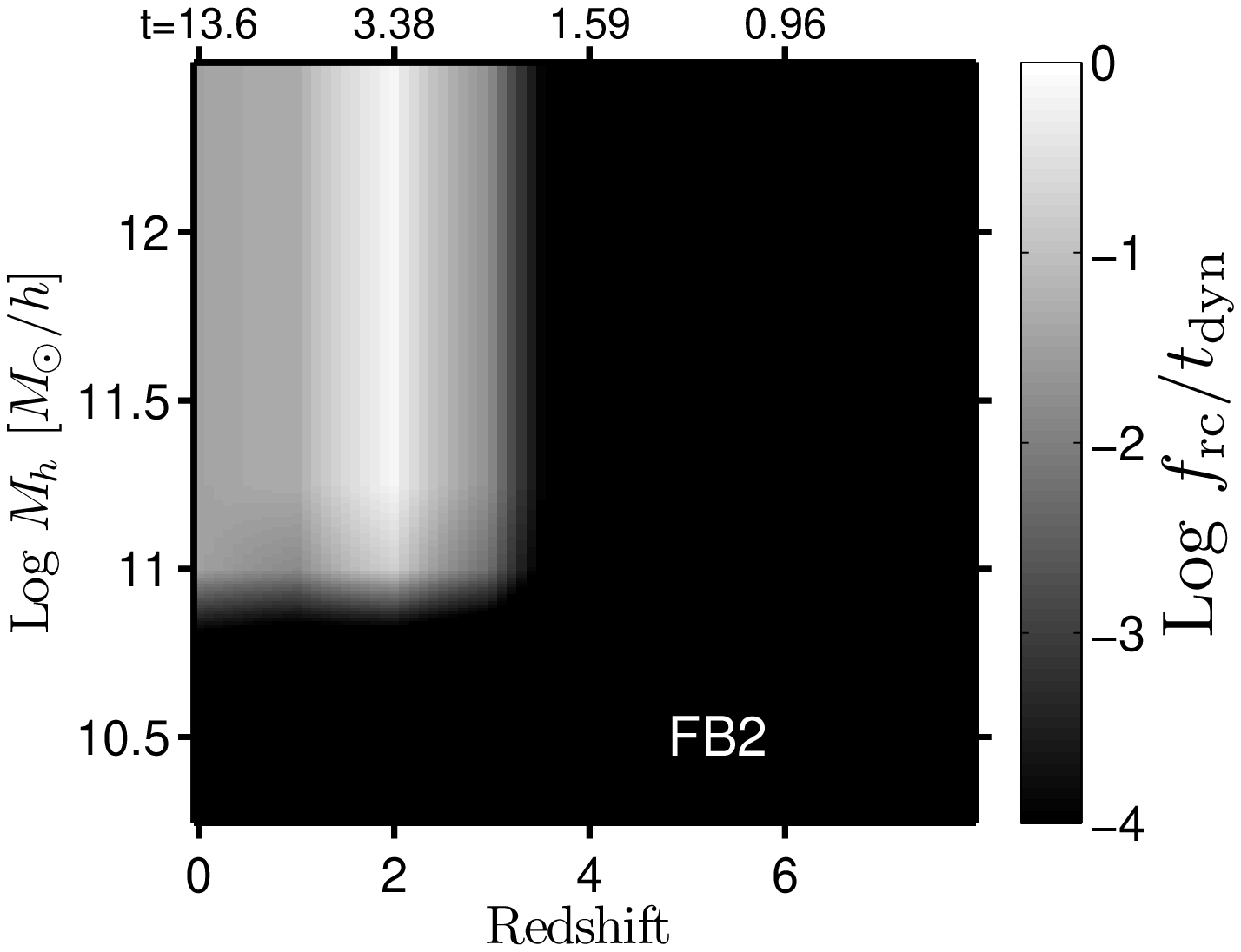,width=220pt}\\
\end{tabular}
\caption{
Feedback as a function of halo mass and redshift in our models.
Top left: feedback efficiency $\ffd$ in the standard model and all SF models.
Top right: $\ffd$ in model FB1.
Bottom left: $\ffd$ in model FB2.
Contours mark log$(\ffd)=$ -0.5, 0, 0.5, 1, 1.5 and 2.
Bottom right: reincorporation efficiency $\frc$ in units of $1/t_{\rm dyn}$
in model FB2.
The main new feature in the FB models is strong feedback at $z\geq 4$,
and especially so in massive galaxies.
Reincorporation is effective at $z<3$ and in massive galaxies.
}
\label{fig:fb_eff}
\end{figure*}

\subsection{The standard model}
\label{sec:standard}

The ingredients
of this model  were chosen based on the simplicity of the physical
processes involved and consistency with observational constraints.

\begin{itemize}
\item  We assume  that all  accretion is  cold, with  $\fc  = 1/t_{\rm
  dyn}$, and that accretion is quenched at $z<3$ above $M_{\rm h}= 1.6
  \cdot 10^{12}  M_{\odot}$ (e.g. Dekel \& Birnboim 2006; Cattaneo et al. 2006;
Ocvirk et al. 2008; Dekel et al. 2009).  To avoid sharp breaks
in the stellar mass function, we smoothed the transition both in
redshift and in mass by hand, as shown in Fig. \ref{fig:cooling}.
\item The rate at which cold gas is turned into stars is 
\begin{equation}
\fs=\es/t_{\rm dyn} \,,
\end{equation}
where  $\es$ is  the star-formation  efficiency.
For $t_{\rm dyn,  disk}$, we assume
\begin{equation}
t_{\rm dyn,  disk} = \frac{3\lambda \cdot  R_{\rm vir}}{\sqrt{2} \cdot
  V_{\rm vir}} \sim 0.0072 \cdot t_{\rm Hubble} \,,
\end{equation}
where the
 halo  spin parameter $\lambda = 0.03$ (according to
the  mean value  found by  Mu\~{n}oz-Cuartas et  al. 2010 in N-body
simulations).
 We  use a
constant star formation efficiency $\es=0.01$, as shown in Fig. \ref{fig:sf_eff} as blue line.
This is comparable to the estimates
of Krumholz \& Tan (2007), and similar to the efficiencies used  in
other recent models (e.g. Krumholz \& Dekel 2010; Bouch\'{e}  et al.  2010  and  Guo  et
al. 2010).
\item  We include  a  simple prescription  for  stellar feedback
which is the usual way to
reproduce the low mass end of the stellar mass function
 (Dekel \& Silk 1986).  The
  formation of stars affects the  surrounding cold gas, making part of
  it unavailable for star formation  for a certain amount of time. The
  rate at which gas is made unavailable for star formation is equal to
  the  SFR times  an efficiency  $\ffd$.  As  in most
  other  SAMs, $\ffd$ is  a combination  of a  constant term  (which is
  called the  ``reheating'' part  of feedback in  De Lucia  \& Blaizot
  2007),  and a  term that  is  inversely proportional  to the  virial
  velocity  to  some power,  which  should  roughly  mimic ejection  of
  material due to  SN explosions (e.g. Bower et al.  2006; De Lucia \&
  Blaizot 2007; Guo et al. 2010).  The feedback
efficiency is thus given by %
\begin{equation}
\ffd = \delta + (\frac{V_{\rm vir}}{\gamma})^{-\alpha}.
\end{equation}
We   use  $\alpha$=3.5,  $\gamma$=161   km/s,  $\delta=1$\footnote{A
non-zero delta  is necessary  in order to  have some feedback  even in
massive galaxies -- otherwise, even in the absence of cooling, the gas
coming from stellar recycling and mergers
 is enough to keep up relatively high star formation
rates in massive galaxies.}.  We do not let log($\ffd$) be higher than 2.5,
and we enforce a constant value for $\ffd$ at $z \geq 5.7$.
  The value of
$\ffd$  as   a  function  of  halo   mass  and  time   is  shown  in
  Fig.~\ref{fig:fb_eff}, top panel.
\item There is no reincorporation, i.e.  we assume that
  cold  gas that  has  been  made unavailable  for  star formation  by
  feedback once never returns back to the cold gas reservoir.
\item  We treat dynamical  friction as  in NW10,  with a  prefactor of
  $\alpha_{\rm df}=$3.
\item No merger-induced star bursts are included.
\item  The  gas  residing in the filaments of satellites  is  stripped with  an  exponential
  timescale of 4 Gyr.
\end{itemize}
Overall our model has similar basic scalings as other SAMs, although
the  parameter values  are  slightly different,  and  the recipes  are
simplified.

We ran  the standard model  (as well as  all following models)  on the
full volume of the Millennium  Simulation (Springel et al. 2005).  The
resulting stellar mass functions (SMFs) at different redshifts and the
sSFR-$\z$ relation (defined in what follows as the sSFR as a function of
redshift at stellar mass $2 \cdot 10^9 - 10^{10}  M_{\odot}$)
 are   shown   in  Fig.~\ref{fig:mass_funs}   and
\ref{fig:ssfrz}  as  thin  blue   lines.
To account  for the observational  completeness limit as  indicated by
Stark et al. (2009), we  only take into account galaxies with log(SFR)
$>$ 0.25, 0.45 and 0.5 $\rm{yr}^{-1}$  at $z$=4, 5 and 6 respectively in
Fig.~\ref{fig:ssfrz}.  Clearly, the observed  relation in the mass bin
$2 \cdot 10^9 - 10^{10}  M_{\odot}$ is not reproduced, very similar to
the  models  from NW10  shown  in  Fig.~\ref{fig:nw10_ssfr}. On the
other hand, note that the the evolution of the SMF is reproduced well
(Fig.\ref{fig:mass_funs}),
which is important for the following discussion of non-standard models.
We show the relation between sSFR and stellar mass at $z$=4 and $z$=6
in Fig. \ref{fig:sSFR_mass4}, top left panels.
The evolution of the mean cosmological SFR density is shown
 as thin blue line in Fig. \ref{fig:res_madau}.
 We  have
checked that  including merger-induced  bursts according to  Croton et
al.  (2006) does  not have  a significant  impact on  any of  the results  shown
here.

\begin{figure*}
\centerline{\psfig{file=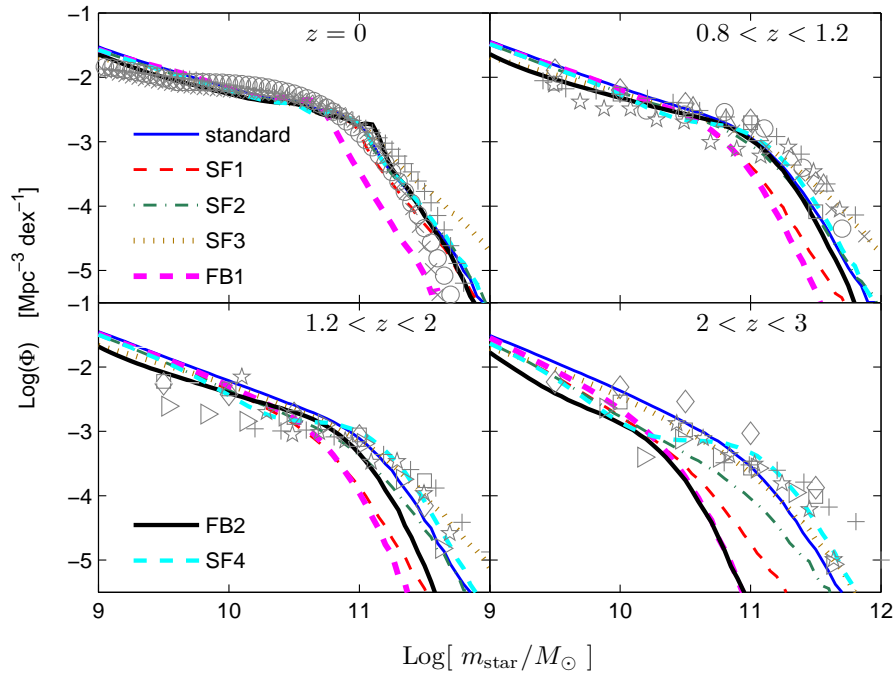,width=130mm,bbllx=30mm,bblly=90mm,bburx=188mm,bbury=210mm,clip=}}
\caption{
Stellar  mass  function  (SMF)  of   galaxies  at  different
  redshift bins, for  our different  models as indicated
  and as summarized in table \ref{tab:models}. Observational
  data are marked by grey symbols.
 At $z>0$, the model stellar mass is convolved with a  Gaussian  error
  distribution of standard  deviation 0.25 dex, which includes the
differences in the IMFs assumed by the different observers.
Data at $z=0$ are by Li et  al. (2009, circles), Baldry et al. (2008,
  crosses), and Panter et  al. (2007, pluses).   Data at $z>0$
are from Bundy  et al.  (2006, $\z=0.75-1$,
  circles),    Borch    et    al.   (2006,    $\z=0.8-1$,    crosses),
  P\'{e}rez-Gonz\'{a}lez   et  al.   (2008,   $\z=0.8-1$,  $\z=1.6-2$,
  $\z=2.5-3$,   plus   signs),   Fontana  et   al.   (2006,$\z=0.8-1$,
  $\z=1.6-2$,  $\z=2-3$,  stars),  Drory  et  al.  (2004,  $\z=0.8-1$,
  upward-pointing  triangles),  Drory  et al.  (2005,  $\z=0.75-1.25$,
  $\z=1.75-2.25$,  $z=2.25-3$ , diamonds  and squares),  Marchesini et
  al.  (2009, $\z=1.3-2$,  $\z=2-3$, right-pointing  triangles).
Model SMFs are  plotted at  $\z=0$  (top left  panel),
  $\z=0.8, 1,  1.2$ (top  right panel), $z=1.2,  1.5, 2$  (bottom left
  panel) and  \z=2, 2.5,  3 (bottom right  panel).
}
\label{fig:mass_funs}
\end{figure*}

 \begin{figure*}
\centerline{\psfig{file=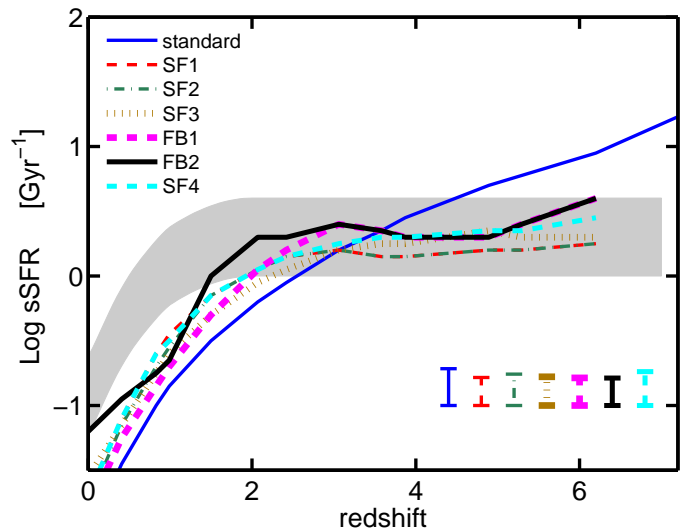,width=100mm}}
\caption{Evolution of sSFR for galaxies with stellar mass in the fixed bin
 $2  \cdot  10^{9}  -  10^{10}  M_{\odot}$ for the models indicated and
listed in table \ref{tab:models}.  At $z \geq 4$, only galaxies with SFR above
  the completeness limits given by Stark et al. (2009) are included.
Errorbars denote $\pm 1\sigma$ at $z>2$.
Each of the redshift bins contains more than 50 galaxies.
All the models shown reproduce the sSFR plateau at $z=2-6$.}
\label{fig:ssfrz}
\end{figure*}

\begin{figure*}
\begin{tabular}{c}
\centerline{\psfig{file=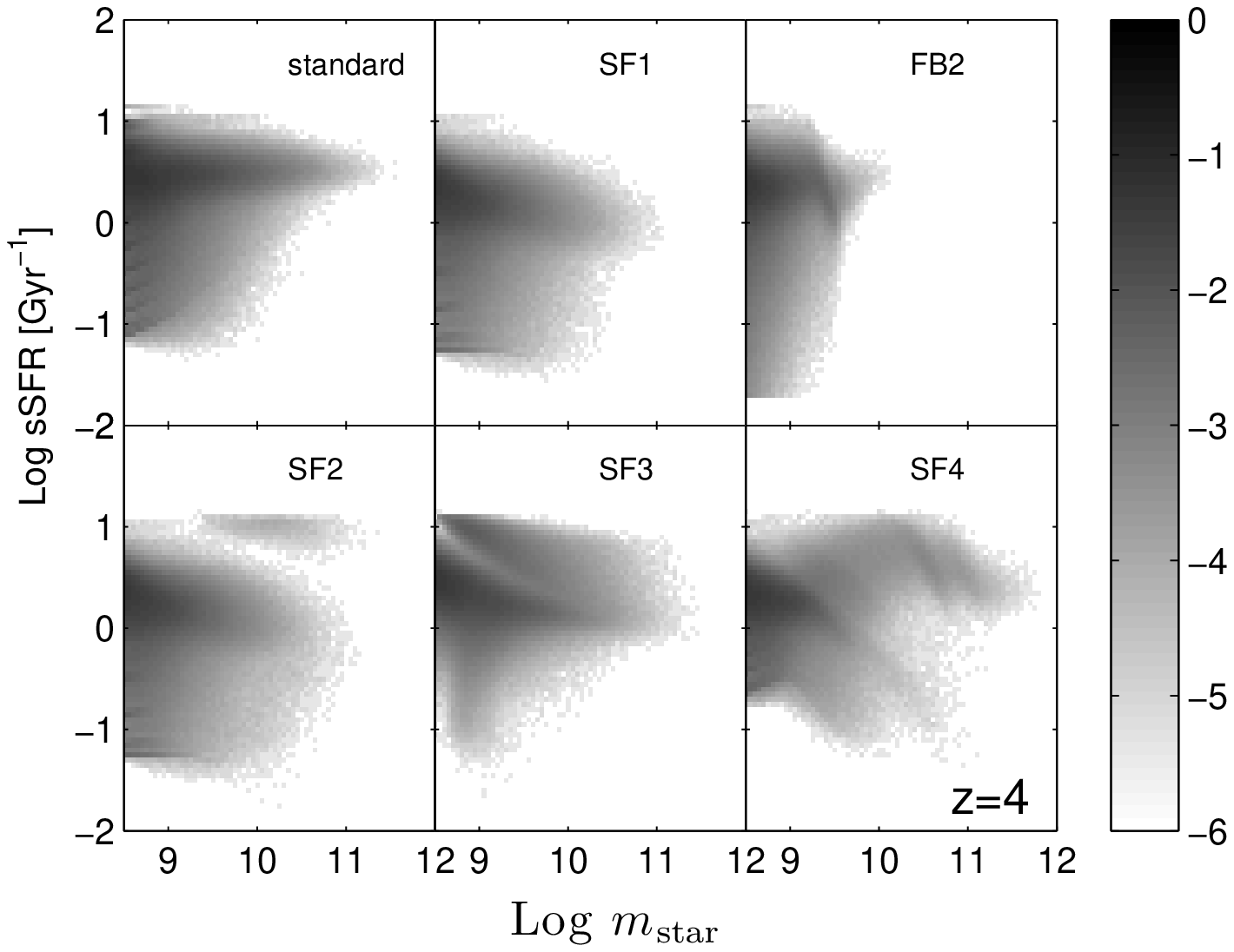,width=300pt}}\\
\centerline{\psfig{file=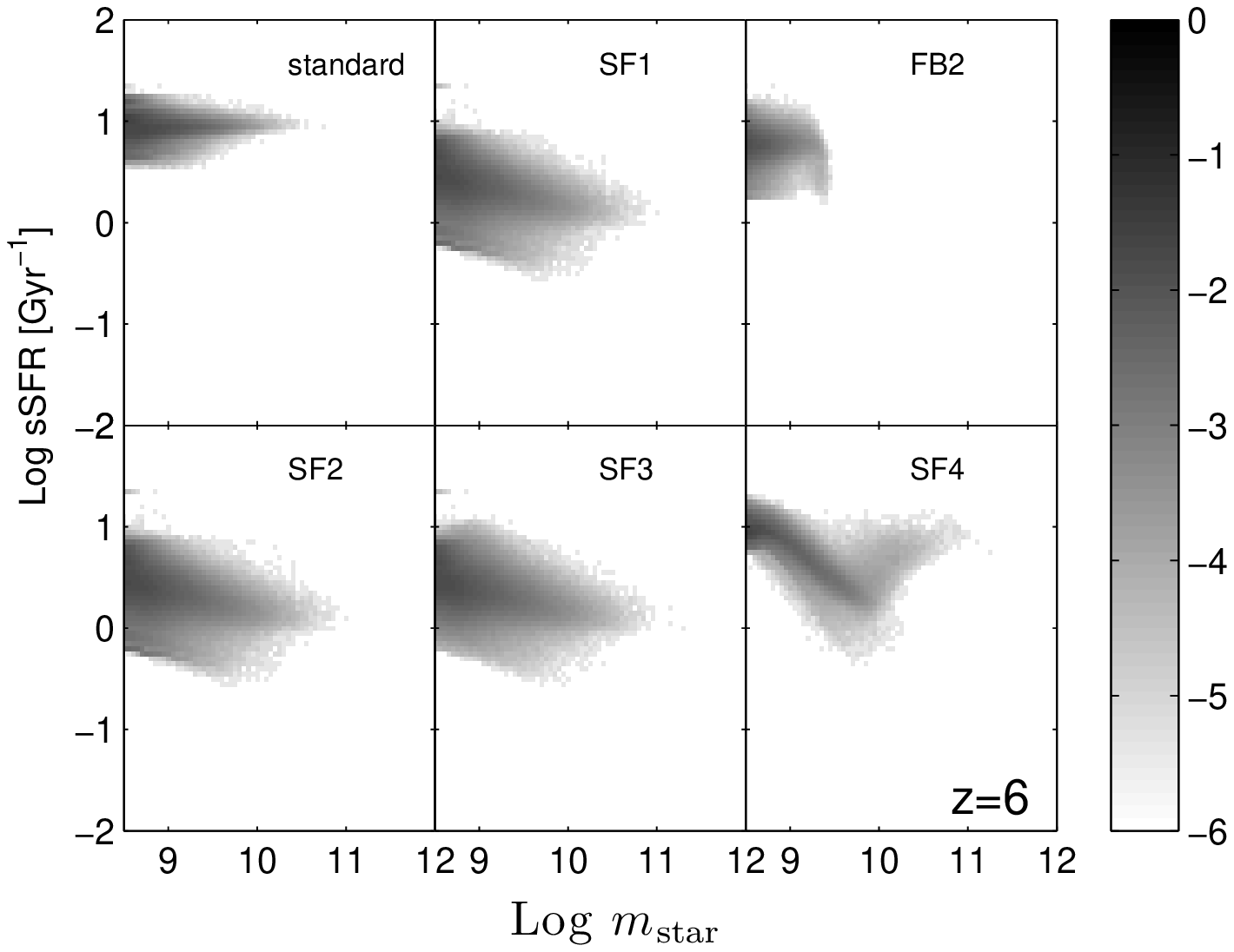,width=300pt}}
\end{tabular}
\caption{
sSFR as a function of stellar mass, at
$z$=4 (top panels) and $z=6$ (bottom panels). The results of model FB1 are
very similar to model FB2.
The greyscale refers to log number density of galaxies in units
of ${\rm Mpc}^{-3}{\rm dex}^{-2}.$}
\label{fig:sSFR_mass4}
\end{figure*}

\subsection{Tuning}
\label{sec:tuning}

We next gradually modify selected ingredients of the standard model.
Our first goal is to bring the median sSFR at $z=2-6$
and in the mass bin $(0.2-1)\cdot 10^{10} M_{\odot}$
into better agreement with the observed plateau,
the grey band in Fig.~\ref{fig:obs}. We also
consider the relation between sSFR and stellar mass at $z=4$ and $z=6$,
shown in Fig. \ref{fig:sSFR_mass4}, in order to verify
that the median sSFR
tends to decrease with increasing stellar mass.
We do not attempt
a fit with the observed sSFR-mass relation
outside the mass bin  $(0.2-1)\cdot 10^{10} M_{\odot}$,
because the lower mass
bin is strongly affected by incompleteness (e.g. Stringer et al. 2011),
while results for the higher mass bin seem controversial
(compare Stark et al. 2009 to Lee et al. 2010a).
Our secondary goal is to avoid severe deviations from the
observed SMF at intermediate redshifts, and from the
observed evolution of mean SFR density.
We do not attempt to fit the observed sSFR at $z<2$. 

This is because we believe the  mismatch in the sSFR between observations and models 
at $z<2$, where the SFR in the universe declines rapidly,
is a separate problem that deserves a dedicated study. It is discussed in more detail
elsewhere (e.g. Fontanot et al. 2009; Karim et al. 2011). Additionally, we find fitting the
$z>2$ sSFR challenging enough even prior to adding
the additional constraints imposed by the low redshift data.
On a separate note, we point out that observations of sSFR in galaxies 
with masses $(0.2-1)\cdot 10^{10} M_{\odot}$  at $z=1-2$ are rather uncertain, probably more
so than at higher redshifts, as they 
tend to fall below the current stellar mass completeness limits at these redshifts
 (e.g.
Rodighiero et al. 2010, Dunne et al. 2009). This is due to the decline of the global
star formation rate density and the increasing
importance of dust, which decrease the intrinsic luminosity
of galaxies at $z<2$. 

For tuning the       models, we focus on one model ingredient at the time, 
i.e. the star formation efficiency or the feedback efficiency.
Using a table with discrete values of the efficiency
in 8 bins in time, and 10 bins in halo mass, we start by fitting the plateau at the highest redshift
and then subsequently continue to lower redshifts.
We try to keep dependencies on halo mass and time  as monotonic as possible, to limit
the number of possible models, and also because monotonic dependencies are easier to 
motivate physically.
Once we have found a solution for the plateau from $\z=2-7$, we compare the model results 
with the other observational 
constraints. Depending on the outcome, 
we discard the model or improve the 
fit to the stellar mass functions and the sSFR-stellar mass relation by making additional changes to the efficiencies.
Part of our models (SF2, SF3, FB2) are then further improved by simple changes to one or two other
model ingredients.
We point out that we 
have not carried out a systematic study covering all the parameter space, 
but a process of trial and error that is geared towards finding successful solutions which fit the constraints
in question.
Fitting the sSFR plateau and the evolution of the
stellar mass function together is not trivial, and a substantial number of 
iterations were needed for each model to 
arrive at the solutions presented below.

The different models are summarized in table \ref{tab:models}.

\begin{table}
\caption{A summary of the models discussed in this work. {$\es$ is the
star formation efficiency,  $\ffd$ the feedback efficiency,
$\frc$ the reincorporation efficiency, $f_{\rm  burst}$ the efficiency
of merger-induced star bursts, and $\alpha_{\rm df}$ is the dynamical
friction prefactor. $m_1$ and $m_2$ is the sum of the stellar and
cold gas mass in the merger progenitors.
If not listed, elements are kept at the standard
model values.}}
\begin{center}
\begin{tabular}{lcccccc}
\hline Models  & Modifications & line  type \\ \hline standard  & -- &
thin blue  \\ SF1 & $\es(t)$
  & thin dashed  red \\ SF2 &  $\es(t)$, $f_{\rm
  burst}(t, m_1/m_2)$  & dotted-dashed  green  \\ SF3  &  $\es(t)$, $f_{\rm  burst} (t, m_1/m_2)$,
$\alpha_{\rm df}(t)$ & dotted brown \\ SF4 & $\es(t, M_{\rm halo}) $ & dashed cyan \\ FB1 &
$\ffd (t, M_{\rm halo})$ & dashed pink \\ FB2 & $\ffd(t, M_{\rm halo})$, $\frc(t, M_{\rm halo})$ & thick black \\ \hline
\end{tabular}
\end{center}
\label{tab:models}
\end{table}

\subsection{Reproducing the sSFR plateau}

In this section, we explore two alternative modifications to the
standard model that aim at reproducing the sSFR plateau:
(i) models ``SF", in which $\es$ is reduced
at high redshifts, and
 (ii) models ``FB", where the feedback efficiency is
enhanced at high redshifts.
Results for all the models are shown in Fig.~\ref{fig:mass_funs},
\ref{fig:ssfrz}, \ref{fig:sSFR_mass4},
 and \ref{fig:res_madau}.

\subsubsection{Model SF1 - tune $\es$}

In model SF1, we tune the SFE parameter, $\es$, while all other model
parameters are kept fixed as in the standard model.
With $\es$ varying in time as shown in Fig.~\ref{fig:sf_eff} (top panel),
the model reproduces the sSFR plateau at $z=2-6$ as shown in
Fig.~\ref{fig:ssfrz}. 
While $\es$ does not need to depend on halo mass, it is not monotonic with
time. In order to have an sSFR plateau starting from redshift $z_{\rm p}$
(chosen here to be $z_{\rm p} \sim 7$), this model requires an earlier epoch
where $\es$ is well above its fiducial value 0.01.
This early star formation is needed in order to produce enough galaxies
of large-enough stellar mass by $z_{\rm p}$, after which the much-lower SFR
adds only slowly to the stellar mass.
At the onset of the plateau, $\es$ has to drop to values well below $0.01$,
and it should continue to gradually decline till $z \sim 4$,
in order to permit the observed low sSFR values at $4<z<z_{\rm p}$.
After $z \sim 4$, $\es$ is gradually increasing in order to match the
high observed sSFR in this regime.  It catches up with the fiducial value
$\es \sim 0.01$ only at low redshifts.
We note that model SF1 predicts that the sSFR plateau does not extend all
the way to the epoch of the emergence of the first galaxies; it is preceded
by a starburst epoch.

Unfortunately, Fig.~\ref{fig:mass_funs} shows that
model SF1 does not produce enough massive galaxies
to match the observed mass function at intermediate redshifts,
despite the high initial $\es$.
Model SF1 thus reveals a
nontrivial tension
between two sets of data, namely the low values of sSFR
at high redshifts
versus the high-mass end of the SMF at intermediate redshifts.
This indicates that the high values of sSFR obtained in SAMs
at high redshifts are needed there for the purpose of building up massive
enough galaxies by $z \sim 2$, and are therefore not easy to avoid.

\begin{figure}
\begin{tabular}{cc}
\psfig{file=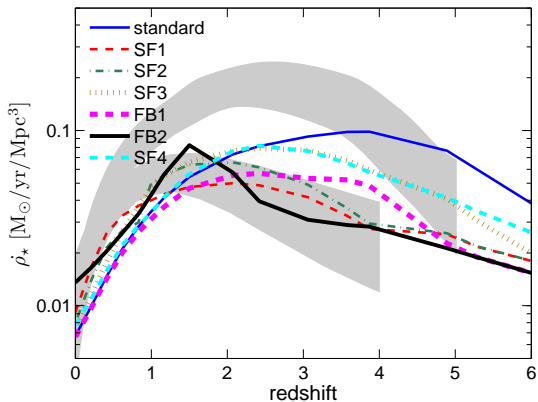,width=220pt} \\
\end{tabular}
\caption{
Cosmological evolution of SFR density for models as indicated
versus observations.
The upper grey belt represents the compilation by Hopkins \& Beacom
(2006, 3$\sigma$ confidence level) from measurements of SFR. More
recent determination by Bouwens et al. (2009) and by Kistler et al. (2009)
from gamma-ray bursts are consistent with the upper grey belt.
The lower grey belt is from  Wilkins et  al. (2008;  1$\sigma$
confidence level), derived from the growth curve of the
stellar mass density.
The model predictions lie in between. 
}
\label{fig:res_madau}
\end{figure}

\subsubsection{Model FB1 - tune feedback}

In model FB1, we only tune the stellar feedback parameter $\ffd$,
while all other model parameters are kept fixed as in the standard model.
The values of $\ffd$ as a function of mass and redshift are shown in
Fig.~\ref{fig:fb_eff} (top-right panel), and the resultant sSFR evolution
and SMF are shown in Fig.~\ref{fig:mass_funs} and \ref{fig:ssfrz}.
We find that fitting the sSFR plateau requires very high
stellar feedback efficiencies at early times ($z \gsim 3$),
and especially so for massive haloes.
This is needed in order to balance the high accretion rates of these haloes.
High feedback efficiency is not needed in the massive haloes at
later times because the cooling in them is set to zero at $z \geq 3$.
At lower
halo  masses $M_{\rm  h} \lsim  10^{11} M_{\odot}$,  moderately strong
feedback is needed at all redshifts. Note that stellar feedback is not
entirely monotonic here  with either mass or time;  we improve this in
the  following model FB2  by adding  reincorporation as  an additional
model ingredient.

As in model SF1, we do not reproduce the high mass end of the SMF with
this  model,  again  indicating  that suppressing  the  efficiency  of
star formation at  high redshift leads to an underproduction
of massive galaxies, if no other changes to the model are made.
As can be seen in Fig. \ref{fig:mass_funs},
this shortage is more severe than in the SF models,
and more so at higher redshifts.
We did not manage to improve this aspect of the FB1 model so far,
since tuning the FB models
is more difficult than tuning the SF models. The reason for this is that
the star formation rate of galaxies is directly proportional
to $\epsilon_s$, while its dependence on the feedback efficiency
is non-linear.

All our models that reproduce the sSFR plateau at $z<  7$ require an
earlier phase of high SFR, with a sSFR higher
than the plateau level by a factor of a few. This provides a high
enough stellar mass at $z \sim 6$,
which allows the desired low sSFR at the plateau level while the SFR
is driven to high values by the high accretion rate.
For example, in order to reach by $z \sim 7$ a mass of $2 \cdot 10^9
M_{\odot}$,
the minimum mass of galaxies on the plateau, a
galaxy needs an average SFR $\sim$ 10 $M_{\odot}{\rm yr}^{-1}$ between
the onset of star formation at $z \sim 9$
and $z \sim 7$, say. This implies a sSFR $\sim 5\, {\rm Gyr}^{-1}$ at
$z \sim 7$, and even higher values
at earlier times, when the stellar mass is smaller.
Such an early phase of high SFR has to be introduced by hand when $
\epsilon_s$ is set to low values at $z<7$,
as in model SF1. 
In model FB1, on the other hand, the high SFR at $z>7$ occurs 
naturally, because the gas available for SFR in a
growing galaxy is proportional to the instantaneous accretion rate,
which is only slowly increasing with time. This is in contrast to the SF 
models, where the SFR is proportional to the
accumulated gas in the galaxy and therefore tends to increase faster 
with time.

\subsubsection{Model FB2 -
tune feedback and reincorporation}

In model FB2,  we improve on model FB1  by adding reincorporation from
the  feedback phase  back to  the cold  phase as  an  additional model
ingredient.  This  makes  it  possible  to keep  the
variation of the feedback efficiency with time monotonic
(although the trend
is of opposite sense at low and high halo masses).
With  model FB2,  we  can
reproduce   the  sSFR   plateau  very   well  down   to  $z   \sim  2$
(Fig.~\ref{fig:ssfrz},  black thick  solid  line), but  again fail  to
simultaneously reproduce the high mass end  of the SMF at $\z \gsim$ 2
(Fig.~\ref{fig:mass_funs},  black  thick  solid line).   Feedback  and
reincorporation   efficiencies    in   model   FB2    are   shown   in
Fig.~\ref{fig:fb_eff},     left     and     right     bottom     panel
respectively. Reincorporation is needed in  order to keep up high sSFR
at  $\z  <  3$  for  haloes  with  masses  $M_{\rm  h}  \gsim  10^{11}
M_{\odot}$.  We set  it to  very low  values outside  this  range. The
efficiency of reincorporation in the  range of halo mass and time when
it is  needed then  quite similar at  all redshifts when  expressed in
units of $1/t_{\rm dyn}$, with a slight peak at around $\z \sim 2$.

\subsection{Reproducing the sSFR and the evolution of the SMF}

While we have found in the previous section that it is
possible to reproduce the sSFR plateau once the
star-formation or feedback efficiency is allowed to vary in time
and with mass,
all the models discussed so far underproduce
the high mass end of the SMF  at $z \gsim 2$.  In this section, we
attempt to improve the fit to the SMF at
intermediate redshifts by an additional modification to the model.
For simplicity, we focus on modifications to model SF1.
Figure \ref{fig:mass_funs} shows that these models, SF2-SF4,
reproduce the high
mass end of the intermediate-redshift SMF better than the previous models.
Figure \ref{fig:sSFR_mass4} demonstrates that this success is associated
with a population of galaxies more massive than $10^9 M_{\odot}$ with high SFR
at $z \sim 4$,
which were missing in the previous models.
We explore here three different ways for producing this population.
In model SF2, we boost the efficiency of
merger-induced star bursts.
In model SF3, we speed up the merger
rate.
In model SF4, we introduce a rather
involved variation of $\epsilon_s$ both with halo mass and time,
making star formation more efficient in relatively high mass haloes while
keeping it inefficient in lower mass haloes.

\subsubsection{Model SF2 - tune $\es$ and mergers}
In model SF2,
we reproduce the required massive galaxies at $z \sim 2$ by boosting up
merger-induced starbursts.
We first tried
the starburst prescription of Croton et  al. (2006)
where $f_{\rm burst}=0.56(m_1/m_2)^{0.7}$, but this had only
a little effect on the SMF.
However, with $f_{\rm  burst}=(m_1/m_2)^{0.3}$ at $z>1$,
followed by the Croton et  al. (2006) prescription at  $z<1$,
model SF2 produces a higher
abundance of high mass galaxies in better
agreement with the SMF at intermediate redshift
(Fig.~\ref{fig:mass_funs}),
while  still   reproducing   the  sSFR   plateau
(Fig.~\ref{fig:ssfrz}).

An additional parameter governing merger-induced starbursts is the burst
duration, which is set to 10 Myr. We find that smearing the bursts over a much longer duration
(e.g. 500 Myr) makes very little difference to the results.
As we only consider the median and the 68\% range, a small population
of starbursting galaxies can be present without ruining the sSFR plateau.
Strongly star bursting galaxies might  also not make it into the Stark
et al.  (2009) or  Gonz\'{a}lez et al.  (2010) samples, as  the highly
star forming regions  in galaxies tend to be  strongly dust attenuated
(e.g Charlot \& Fall 2000), and are therefore a good way
of building up stellar mass in a hidden mode.

\subsubsection{Model SF3 - tune $\es$ and mergers}

Model SF3  is a  variation of  model SF2, where the high starburst
efficiency at $\z>1$ is replaced by a shorter characteristic time for dynamical
friction. The fiducial dynamical friction prefactor of
$\alpha_{\rm df}=3$ is replaced by
$\alpha_{\rm df}=0.1$ at $z>1$ and $\alpha_{\rm  df}=5$ at $z<1$,
while the starbursts are moderate, $f_{\rm  burst}=0.2(m_1/m_2)^{0.7}$.
The rapid stellar assembly
in this model boosts up the buildup of stellar mass, and we need to lower the efficiency
of star formation accordingly (Fig.~\ref{fig:sf_eff}, top panel).
This model provides a sensible fit to the SMF (Fig.~\ref{fig:mass_funs})
and it reproduces the sSFR plateau (Fig.~\ref{fig:ssfrz}).

\subsubsection{Model SF4 - tune $\es$}

Model SF4 demonstrates a third way to build up high mass
galaxies by moderate redshifts without enhancing the contribution of mergers.
In this model, the star-formation efficiency, $\epsilon_s$, is varied
as a function of both halo mass and time as shown in Fig. \ref{fig:sf_eff}.
This allows for a simultaneous fit to the SMF (Fig.~\ref{fig:mass_funs})
and the sSFR plateau (Fig.~\ref{fig:ssfrz}).
We note that in this model, the sSFR does not in general decrease with 
increasing stellar mass, as shown in Fig. \ref{fig:sSFR_mass4}, for the z=6
results, bottom right panel. This is contrary to all observations of
this relation we are aware of, which tend to find that the median sSFR of galaxies
decreases with increasing stellar mass.

\subsection{Summary of model results}
In this section, we summarize the successes and failures of the models, 
guided by Fig. ~\ref{fig:mass_funs}, \ref{fig:ssfrz} and ~\ref{fig:res_madau}.
From Fig. \ref{fig:mass_funs}, which shows the stellar mass functions
up to $z \sim 3$, it becomes clear that 
only models SF3 and SF4 manage to reproduce the SMF at $z>1.2$.

In Fig. \ref{fig:ssfrz}, we show that all our models except the
standard models provide a reasonable fit to the sSFR plateau from $z \sim 2-6$. 
At $z<2$, the sSFR is lower than the observational estimates.
 This may be partially
due to incompleteness in the observations, which misses galaxies
with low SFR especially at low redshift, where the 
global SF in the Universe has declined. The mismatch at $z \sim 0$, however,
is probably real and already apparent in the standard model. It is likely 
connected to a similar underproduction of the sSFR present in 
other recent SAMs (e.g. Fontanot et al. 2009, Guo et al. 2011). We note
that our efforts to improve the match to the sSFR plateau lead to an 
increase in the 
sSFR at $z<2$ and thus to a better agreement with observations.

In Fig. \ref{fig:res_madau} we show the global star formation 
rate density as a function redshift in our models. The upper grey belt represents 
the directly observed SFR density from Hopkins \& Beacom (2006),
 while the lower belt is derived
from the growth  curve of the stellar mass (Wilkins et al. 2008). 
More recent estimates like from Bouwens et al. (2009) and
from Kistler et al. (2009) are in agreement with the upper belt.
The observational results that are directly measured from star formation, 
and those that are obtained
indirectly from the
evolution of the stellar mass density 
are thus clearly inconsistent, which is the reason
that SAMs (like e.g. Guo et al. 2011) usually have a SF density somewhere in between
these two constraints. 
Results of our
various models start to deviate considerably at $z>2$, but the discrepancy between
the observational results makes it impossible to use them to constrain models.
 The highest global SFR density
at $z>3$  is reached by the standard model, followed
by the models which have the best fit to the SMF at high redshift, 
namely model SF3 and SF4.

\subsection{The sSFR of individual galaxies}
\label{sec:individual}

\begin{figure*}
\centerline{\psfig{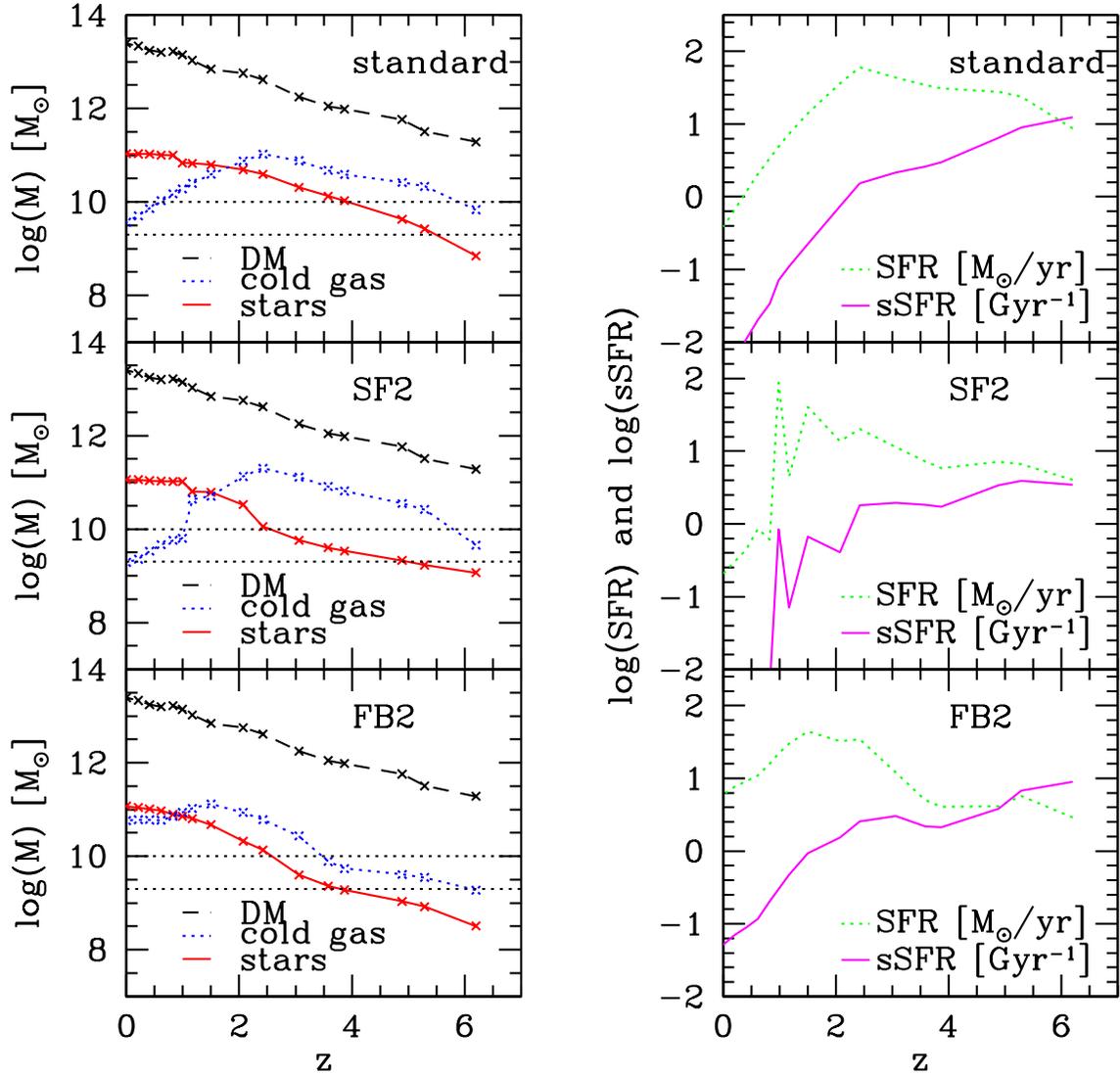}}
\caption{
Evolution in a single galaxy within the same halo according to different
models.
Top row: standard model.
Middle row: model SF2.
Bottom row: model FB2.
Left: mass in dark matter (same in all models), stellar mass, and cold gas.
Right: SFR and sSFR.
Horizontal dashed lines indicate the stellar mass range
relevant for the median sSFR shown in Fig.~\ref{fig:ssfrz}.
The peaks in the SF history of model SF2 are due to starbursts.
}
\label{fig:connect}
\end{figure*}

In Fig.~\ref{fig:connect}, we follow the growth of one individual
model galaxy following the main branch of its merger history.
Shown as a function of redshift is the mass in the dark-matter, cold gas and
stars, the SFR and sSFR, for the standard model, SF2 and FB2.
The resulting stellar  mass at $z=0$ is similar in the three models.
However, the star-formation history and the growth of stellar mass
are different --- in both model SF2 and FB2, the buildup of stellar mass
is delayed compared to the standard model.

We note that at $z>2$ the SFR in individual galaxies
is rising with time for all models.
This is a common feature to all our model galaxies, and is consistent with
the predictions
 of Finlator et al. (2006, 2011), as well as with the observational
finding of Papovich  et al. (2010), Maraston et al. (2010),
Lee et al. (2010a,b).
It is emerging naturally from the fact that in an individual galaxy, 
 the accretion rate is expected 
to be growing in time in its rapid growth phase at $z \geq 2$ (Neistein
\& Dekel 2008).

We see in Fig. \ref{fig:connect}
that as the galaxy evolves from $z=6$ to $z=2$,
the sSFR declines in a rather slow pace under models SF2 and FB2
compared to the steeper decline in the standard model.
The sSFR plateau discussed in the previous sections, which refers to the
median value for galaxies of a fixed mass at different redshifts,
does not necessarily imply a constant sSFR in individual galaxies as
they grow. However, the observations of Stark et  al. (2009) and Papovich
et al.  (2010, Figs. 2 and 3) do argue for a constant sSFR in individual
galaxies as well. It is thus encouraging that our models that reproduce
the sSFR plateau also come close to reproducing the constancy of the sSFR
in individual galaxies.

\section{Discussion}
\label{sec:discuss}
In subsections \ref{sec:change_quiescent} and \ref{sec:change_mergers}
of this discussion section, we comment on the physical plausibility
of the ad-hoc modifications suggested by our models.
This discussion is not complete or conclusive --- it is only meant
to raise some of the relevant issues and to trigger further discussion.
The limited goal of the current paper remains to point out the possible
nature of the modifications to the standard models that are required
for reproducing the sSFR plateau together with the SMF at moderate redshift.

\subsection{Changes to the quiescent evolution of galaxies}
\label{sec:change_quiescent}

\begin{enumerate}
\item \emph {Star  formation efficiency:}
In the SF models, $\es$ has to vary in time, with a very high
efficiency at $z \geq 7$, and a low efficiency at $z =  3-6$. The latter
is consistent with the prediction of Gnedin \& Kravtsov (2010) that the
low metallicity and high UV radiation at $z \sim 3$ should significantly
lower the normalization of the Kennicutt-Schmidt relation.
This effect is particularly strong for low gas surface densities. 
An effective suppression of star formation in high redshift galaxies and at low masses turns out 
to be a natural outcome of the low metallicity in these galaxies (Krumholz \& Dekel 2011). 
The low dust content enables the UV radiation from stars to heat the gas (while dissociating H$_{2}$ molecules) 
and to prevent further star formation, until enough metallicity is built up and the SFR regions are shielded.
Observational studies at $z \sim$ 3 seem to agree with this
  prediction (e.g. Wolfe \& Chen 2006; Rafelski et al. 2010).
Furthermore, both SAMs and hydrodynamical simulations have suggested
that a low $\es$ at high redshift improves the match with observations
in a number of ways (Baugh et al. 2005; Lacey et al. 2010; Agertz et al. 2011).
It is more difficult to explain the high $\es$ at $z \geq  7$,
necessary for producing enough galaxies with a high stellar mass by
$z \sim 6$. What might help is a feedback mechanism that causes an
abrupt change in the mode of star formation after a very active initial phase.
Model SF4 suggests in addition an increase of $\es$ with halo mass.
Such a variation might result from faster metal enrichment or less efficient
feedback in more massive haloes.

\item \emph {Stellar feedback:}
The FB models suggest strong early stellar feedback that is not limited to
small galaxies.  This is in potential disagreement with conventional models
of SN feedback, in which the effect of feedback at a fixed halo mass is
expected to be weaker at high redshift, where the deeper potential
well makes it harder to eject gas from the galaxy (Dekel \& Silk 1986;
Dekel \& Birnboim 2006).
One should therefore consider additional feedback mechanisms that may
vary in the required way with time and mass.
For  example, a metallicity-dependent stellar feedback (Nishi \& Tashiro 2000; 
Krumholz
\& Dekel 2011)
will be especially effective at high
  redshifts.
The required strong feedback at high halo masses ($\log(M_{\rm h})>11.5$
at $t_{\rm Hubble}  < 2$) may be achieved by AGN feedback from massive
black holes rather than stellar feedback.

\item   \emph{Reincorporation:}
Model FB2 requires a high rate of reincorporation only at $z \lsim  3$.
In other SAMs, the reincorporation rates scale with $1/t_{\rm  dyn}$,
and therefore gradually increase towards higher redshift, making the effective
stellar feedback less efficient at high $z$.
A weaker time dependence, more consistent with FB2,
is found in simulations that incorporate
strong momentum-driven winds (Oppenheimer et al. 2010).
Another option is ``reincorporation" that utilizes a gas reservoir
in the galaxy rather than gas that has been ejected earlier. 
For example, this occurs naturally for metallicity-dependent feedback.
As long as the metal content in a galaxy is low, gas is not well shielded
from UV radiation by dust, leading to efficient heating and $H_2$ dissociation.
Once the metal content reaches a certain threshold value, the gas is shielded and star formation
becomes efficient. 
The abrupt appearance at $z \sim 3$ of reincorporation in FB2 is
an oversimplification, but the general increase of reincorporation
with halo mass is plausible both for ejective and non-ejective feedback.

\end{enumerate}

\subsection{Changes to the recipes for mergers}
\label{sec:change_mergers}

\begin{enumerate}
\item  \emph{Enhanced starburst efficiency at high redshift:}
In model SF2, the starburst efficiency, defined according
to eq. \ref{eq:sf_burst}, is higher at higher \z. 
This is not to be confused with the SF efficiency $\epsilon_s$,
 which is relevant for quiescent star formation.
An increased star burst efficiency at high $\z$ 
is plausible due to the shorter dynamical times. 
The gas-rich mergers at high redshift may in principle be very different
from today's more familiar gas-poor mergers. High-resolution gas-rich merger
simulations indeed indicate an effective star formation efficiency
that is higher than in non-merger situations (Teyssier, Chapon \&
Bournaud 2010).

\item  \emph{Enhanced merger rate at high  redshift:}
In model SF3, the dynamical friction prefactor $\alpha_{\rm df}$ is reduced
to 0.1 at $z>1$, which speeds up the merger process by a factor $\sim 20-30$
compared to the standard model and other  SAMs (Croton  et  al. 2006;
De Lucia  et al. 2007).  At high redshift, the merging galaxies flow from
the virial radius to the central galaxy along narrow radial streams
associated with the cosmic-web filaments in about one half of a halo
crossing time, $R_{\rm vir}/V_{\rm vir}$, which is $\sim 0.1 t_{\rm Hubble}$
(Dekel et al. 2009). This implies a merging time substantially shorter
then the dynamical friction time estimated for gradual spiraling in.
Hopkins et al. (2010) point out that the standard estimate of merger
time based on dynamical friction is indeed an overestimate
when the approach of the satellite is along a radial orbit,
or when the dynamical friction estimate starts at a distance larger
than 0.1 - 0.2 $R_{\rm vir}$, where the satellite is no longer properly
resolved in the simulation.  Both of these conditions tend to be valid at
  high redshift (Wetzel 2010, Hopkins et al. 2010), and together they may
lead to an order of magnitude overestimate of the merger time in the standard
model.  We used the Millennium simulation to verify that with
$\alpha_{\rm df}=0.1$, the merging time is comparable to the halo crossing
time.

\end{enumerate}

\subsection{Dust at high redshift}
\label{sec:dust}

The validity of the sSFR plateau at high redshifts crucially depends
on the correction adopted for dust extinction by the different authors.
At $z \sim 2$, the estimates for dust extinction in LBGs from the UV 
slopes are confirmed by radio estimates (Pannella et al. 2009) and
from comparison with a local sample of Lyman break analogs 
(Overzier et al. (2011). At higher redshifts, the estimates
are more uncertain.
Stark  et al. (2009), Gonz\'{a}lez et al. (2010)
and Labb\'{e}  et al. (2010a,b) all assume practically
no dust extinction at $z>4$.   This  assumption is
supported by simple theoretical considerations and several
observational studies.  It is expected that high-redshift galaxies should
have a lower dust content than today's galaxies simply because they had
less time to produce metals and dust.  Indeed, theoretical models
(e.g. Guo \&  White 2009) often  assume a lower dust
extinction at higher redshift.  Bouwens et  al. (2009) estimated
the dust extinction at high redshift based on the observed UV
continuum slope, and found low values at $z \gsim  4$, especially for
the dominant population of galaxies with relatively low UV luminosities.
Labb\'{e} et al.  (2010b), Finkelstein et al. (2010) and
Bouwens et  al. (2010) also report  on low or zero  dust extinction at
$\z \sim 6-8$.  Brammer \& van Dokkum (2007)  tested whether the usual
LBG selection criteria miss a  significant number of dusty galaxies at
$\z \sim 4$ by selecting according to Balmer breaks instead, and found
that  this  is  not  the  case,  again  indicating  that  strong  dust
obscuration at high redshift is rare. This may however not
be the case for massive galaxies, which potentially already have
high dust content at $z \sim 3-4$ (Mancini et al. 2009, Marchesini et al. 2011)
and thus may be missed by the usual LBG selection criteria.
Also,
 Schaerer \& de  Barros (2010) and Yabe et
al. (2009) argue for high correction factors of $\sim$ 9 when translating
the UV flux to SFR at $z=5-7$. We note that in order to reconcile the
plateau  in the sSFR with  current  SAMs by the effect of dust-extinction
alone, the  dust
extinction would have  to increase with redshift, a  trend which is
opposite to basic theoretical  expectations.
While the ``dust" has not settled yet on this debate,
it seems that the evidence for a sSFR plateau is intriguing enough
to justify a serious theoretical consideration, but it is left for
future observations to tell whether this is indeed a valid strong constraint
or a fluke.

\subsection{Comparison to previous work}
\label{sec:previous_work}

Previous efforts to understand the properties of star forming galaxies
at  $z>4$ include Finlator  et al.   (2006, 2011),
Night et al. (2006), Mao et  al. (2007), Nagamine et al. (2008), Stark
et al.  (2009), Lee et al.  (2009), Khochfar \& Silk  (2011), Lacey et
al. (2010),  Lo Faro et al.  (2009), and Stringer et al. (2010).
Here we compare  our findings to
some  of those  studies.

Lee et  al. (2009) concluded based on clustering analysis that the duty
cycle of star formation in high-redshift galaxies must be short, consisting
of episodes of $<0.35$ Gyr in each galaxy.
Stark et al. (2009) came to similar conclusions.
This would be in apparent conflict with the need to build up the massive
end of the SMF by $z \sim 2-3$.
Our model SF1 that reproduces the sSFR plateau assumes a continuous mode of
star formation, and even then, it underpredicts the massive end of the SMF.
If the low sSFR were associated instead with short-lived
bursts of star formation, it would have been even harder to build up the
stellar mass quickly enough.  A  way  to  reconcile a short duty cycle
with the required fast build-up of stellar mass might be that the short-lived
episodes of SF are only the final phase of a much more active
and dust-obscured starbursts, enough for building up the high mass end of
the SMF but undetectable in their violent phase.
Finlator et al. (2011) suggest that the clustering data by Lee  et al. (2009)
can also be  explained without a bursty star formation history,  if there
are strong outflows that lead  to an increased scatter in the relation
between baryonic mass and host halo mass.

Khochfar \&  Silk (2011) present a  model that reproduces the
sSFR  plateau  at  $z>4$  with  (i) a  very  high
frequency of star-bursts at high  redshift and (ii) a burst efficiency
that  scales with the  inverse of  the halo  circular velocity  to the
third  power.
In this  way, star  formation in galaxies in  the mass
range  relevant  for the  plateau  is  inefficient  at high  redshift, but stellar mass
grows efficiently during mergers in low mass haloes.
Their model thus resembles a combination of our models SF2 and SF3.

The simultaneous need for a low sSFR at high redshift and
a fast build-up of stellar mass seems to be in odds
with the notion that the observed growth rate in stellar mass
density is low compared to that predicted by integrating
the observed SFR density (Wilkins et al. 2008).
These seemingly inconsistent problems may reflect a difference between
the overall evolution of stellar mass density and the evolution of
the population dealt with here, involving relatively massive galaxies in
a fixed mass bin.

Lo Faro et  al. (2009), comparing a SAM for LBGs at $z \sim 4-6$ to
observations, found that their model produces an excess of star-forming
galaxies with stellar masses $\sim 10^{9} - 10^{10} M_{\odot}$ at
$z  \sim 4$, within the range relevant for our study of the sSFR plateau.
They conclude that some feedback mechanism must suppress star formation
at early times in the corresponding haloes of
$\sim  10^{11} -10^{12} M_{\odot}$,
but it should become ineffective at later times in haloes
of similar mass. As Lo Faro  et al. (2009) point out, such a behaviour of the feedback
efficiency is not part of the standard feedback models.
Their conclusions are thus
 similar to ours, albeit they originate from a
different observational constraint.

\section{Conclusions}
\label{sec:conclusions}

We explored possible ad-hoc modifications to standard semi-analytic models
that reproduce a constant sSFR at $z>2$ while growing enough massive
galaxies by $z \sim 2$, and came to the following conclusions.

\begin{itemize}
\item
We confirm that a sSFR plateau at $z=2-6$ is in robust disagreement with
current models, as claimed in Bouch\'{e} et al. (2010).
We have demonstrated that, in a fixed stellar mass bin of
$2\cdot 10^9-10^{10} M_{\odot}$, the common feature of a variety of standard
SAMs is a gradual decline of the sSFR with time, associated
with the decline of the total specific accretion rate.
\item
We find that it is possible to reproduce the sSFR plateau together
with the stellar mass function at $z \sim 2$ via non-trivial modifications
to the  standard SAMs.  Three different modifications seem necessary,
related to three different observational features:
(1) the low sSFR at $z>4$, (2) the high sSFR at $z \sim 2-3$, and (3)
the abundance of high mass galaxies at $z \sim 2-3$.
The low sSFR at $z>4$ is reproduced either by strong stellar feedback at high
redshift in all masses, or by inefficient star formation at high redshift
following a phase of very efficient star formation at very high redshift.
The high sSFR at $z=2-3$ could emerge either by a drop
in the feedback efficiency at this epoch, or by a corresponding enhancement of
star formation efficiency, or by efficient reincorporation of gas
that was previously prevented from forming stars.
Finally, the high mass end of the SMF at $z > 1$ can be generated
despite the low SFR at high redshift by an additional modification of
the star formation in a sub-population of massive galaxies at high redshift.
This is achieved in our models by either speeding up the mergers, or enhancing
the merger-induced starbursts, or by introducing a non-trivial
dependence of star formation efficiency on halo mass.
However, none  of our modified models matches  the stellar
mass function at $\z \sim 3$ as well as our standard model. This
reflects the fact that the low sSFR at high redshift
and the presence of massive galaxies by $\z \sim 3$ are not easily
reconciled.

\item
Our models predict that the SFR in individual galaxies
is monotonically increasing with time at $z >2$.
This is in agreement  with the theoretical predictions of
Finlator  et  al. (2006,  2011)  and  Bouch\'{e} et al. (2010), as well as
the observational results by Papovich  et al. (2010).
We note in particular that an SFR that grows exponentially with time
implies a constant sSFR. It should be mentioned that
the decreasing SFR sometimes assumed in SED fitting
(e.g. in Stark et al. 2009) is incorrect.
\end{itemize}

We have demonstrated that the simple SAM of NW10 is useful for
  exploring how current SAMs should be modified in order to
  match new observational  constraints, and for pointing out
  apparently conflicting observational constraints.  A similar method will be useful in addressing
other puzzling observations such as
the  tilt in the relation of sSFR and stellar  mass
  (e.g. Somerville et al. 2008), the fraction of passive galaxies
as a function of
  stellar mass at $z=0$ (e.g. Weinmann et al. 2010), or the fraction of
  AGNs as a function of stellar mass (Fontanot et al. 2010).

We learn that the observed sSFR at high redshift has the potential for
posing powerful constraints on the physical processes of star formation and
feedback. If the sSFR plateau, as observed by Stark et al. (2009),
Labb\'{e} et al. (2010a, 2010b), and Gonz\'{a}lez et al. (2010), is
confirmed, it will provide invaluable information on the baryonic physical
processes at high $z$, indicating that it could be different from those
at low redshift. One should be eagerly waiting for new developments in the
observations of SFR and stellar mass at high redshift.

While we started a discussion of the physical plausibility of the
required modifications to the standard recipes of galaxy-formation models,
the modifications suggested above are primarily of an ad-hoc nature.
They deserve a thorough theoretical study of physical mechanisms that could
be responsible for the required variation with time and mass.

\section*{Acknowledgments}
We thank Rychard Bouwens, Sadegh Khochfar and the anonymous referee
for helpful comments on the draft,
and Rachel Somerville, Ben Oppenheimer, Ivo Labb\'{e}, Niv Drory,
Emmanuele  Daddi,  Marcel Haas, Raanan  Nordon  and Giulia  Rodighiero
for  useful discussion. EN was partially supported by the Minerva fellowship
during this project.
AD was partially  supported by ISF grant 6/08,
by GIF grant G-1052-104.7/2009,
by a DIP grant, and by NSF grant AST-1010033.
SQL databases containing the Millennium simulations are publicly released at
    \texttt{http://www.mpa-garching.mpg.de/millennium}. The code used to
generate semi-analytic models based on the NW10 method
is publicly available under
\texttt{http://www.mpa-garching.mpg.de/galform/sesam}
The Millennium site was created as part of the activities of the German
    Astrophysical Virtual Observatory.

\clearpage
\end{document}